\documentclass[fleqn,useAMS,usenatbib]{mn2e}
\pdfoutput=1

\usepackage{graphicx}
\usepackage{amsmath}
\usepackage{amssymb}
\usepackage{rotating}
\usepackage{pdflscape}
\usepackage{afterpage}

\newcommand{\rln}{\mathrm{ln}}

\title[PDF tails]{Power-law tails in probability density functions of molecular cloud column density}
\author[C. M. Brunt]{C. M. Brunt$^{1}$\thanks{E-mail brunt@astro.ex.ac.uk} \\
$^{1}$Astrophysics Group, School of Physics, University of Exeter, Stocker Road, Exeter, EX4 4QL, UK\\
}
\begin{document}

\date{Accepted ; Received ; in original form }

\pagerange{\pageref{firstpage}--\pageref{lastpage}} \pubyear{2014}

\maketitle

\label{firstpage}

\begin{abstract}
Power-law tails are often seen in probability density functions (PDFs) 
of molecular cloud column densities, and have been attributed to the effect of gravity.
We show that extinction PDFs of a sample of five molecular clouds obtained at a few tenths of a parsec resolution,
probing extinctions up to A$_{{\mathrm{V}}}$~$\sim$~10~magnitudes, are very well described by lognormal
functions provided that the field selection is tightly constrained to the cold, molecular zone and that
noise and foreground contamination are appropriately accounted for. In general, field selections that
incorporate warm, diffuse material in addition to the cold, molecular material will display apparent core+tail PDFs.
The apparent tail, however, is best understood as the high extinction part of a lognormal PDF arising from
the cold, molecular part of the cloud. We also describe the effects of noise and foreground/background
contamination on the PDF structure, and show that these can, if not appropriately accounted for, induce
spurious tails or amplify any that are truly present. 
\end{abstract}


\section{Introduction}

There has been a great deal of interest lately in the column density probability density functions (PDFs) of molecular clouds.
The PDFs appear lognormal in form, but often with $\sim$power-law tails in the extreme positive wing.
This has been linked to the role of gravity in star-forming regions 
(Kainulainen et al 2009; Froebrich \& Rowles 2010; Russeil et al 2013; Alves de Oliveira et al 2014; Schneider et al 2014), 
as corresponding tails are seen in density PDFs derived from numerical simulations of
isothermal clouds in 3D (Klessen 2001; Cho \& Kim 2011; Girichidis et al 2014).

In this paper we show that fitting the column density PDF in the direct space representation, where the
role of noise is more straightforwardly accounted for, is more reliable than the methods of fitting in 
log space that have generally been used to-date. 
We show that the amplitude (or existence) of power-law tails in column density PDFs can
be significantly affected by noise, and the presence of an unaccounted-for foreground/background 
contribution to column density. 

We also investigate the role of {\it field selection}, paying attention to the relative amounts of cold molecular
gas and warm diffuse (atomic and molecular) gas in the field.
The extinction-based molecular cloud column density fields analyzed below can be fitted with a single lognormal function, provided
that the field is restricted to the cold molecular parts of the extinction and that noise is correctly accounted for in the fitting.
Our results apply to PDFs sampled at a few tenths of a parsec resolution, comprising extinctions up to $\sim$~10~magnitudes.

The layout of the paper is as follows.
In Section~2, we describe the basic idea behind the fitting method. Section~3 and Section~4 describe fitting PDFs with
the standard log-space method and our direct space method, respectively. In Section~5, the systematics of noise and
contamination are described. Section~6 documents the effect of field selection on the derived PDFs, followed by a Discussion
in Section~7. Section~8 includes a brief aside on the scale- and resolution-dependence of derived PDF dispersions. Our Summary
is given in Section~9.

\begin{figure*}
\includegraphics[width=54mm]{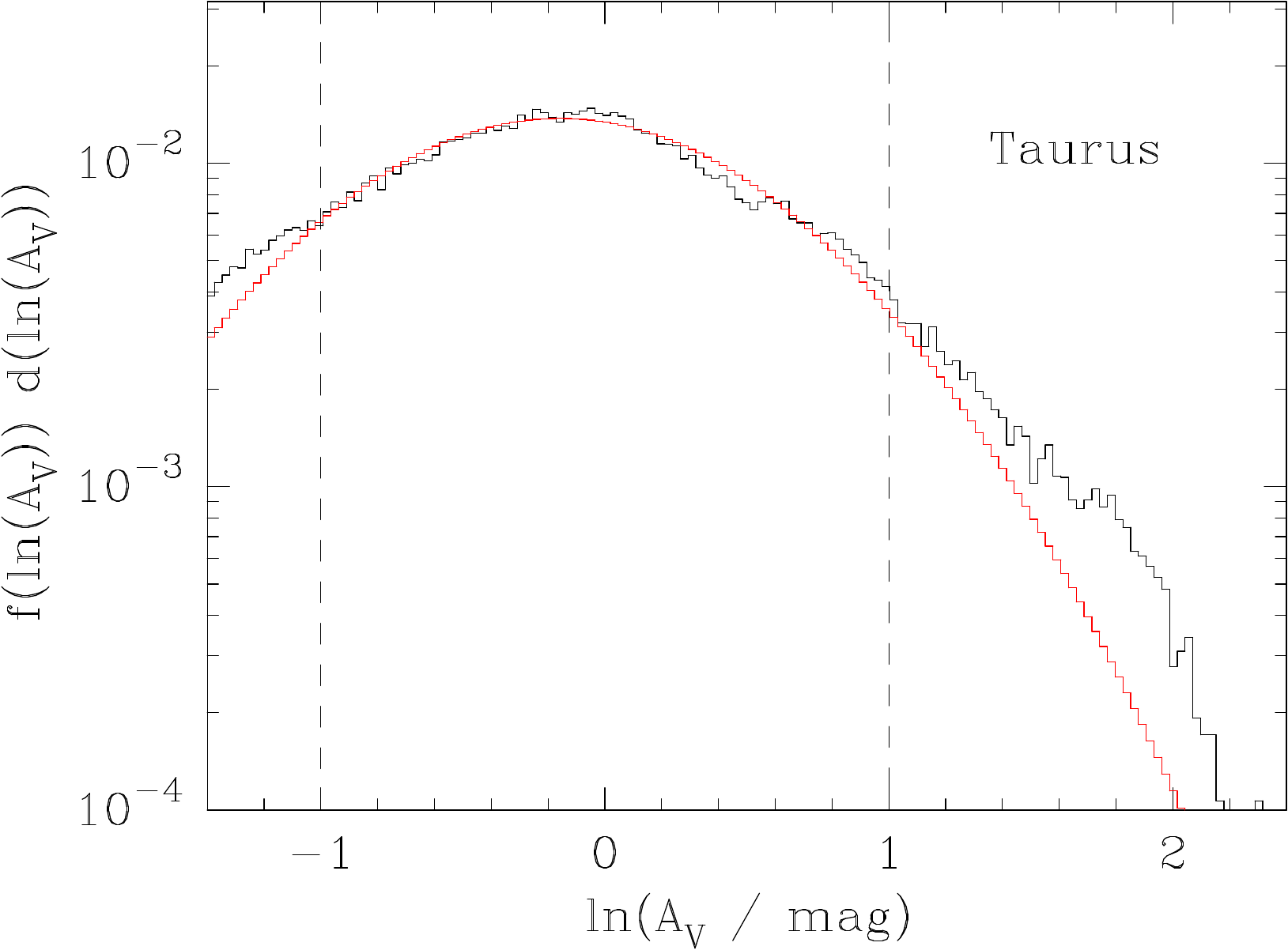}
\includegraphics[width=54mm]{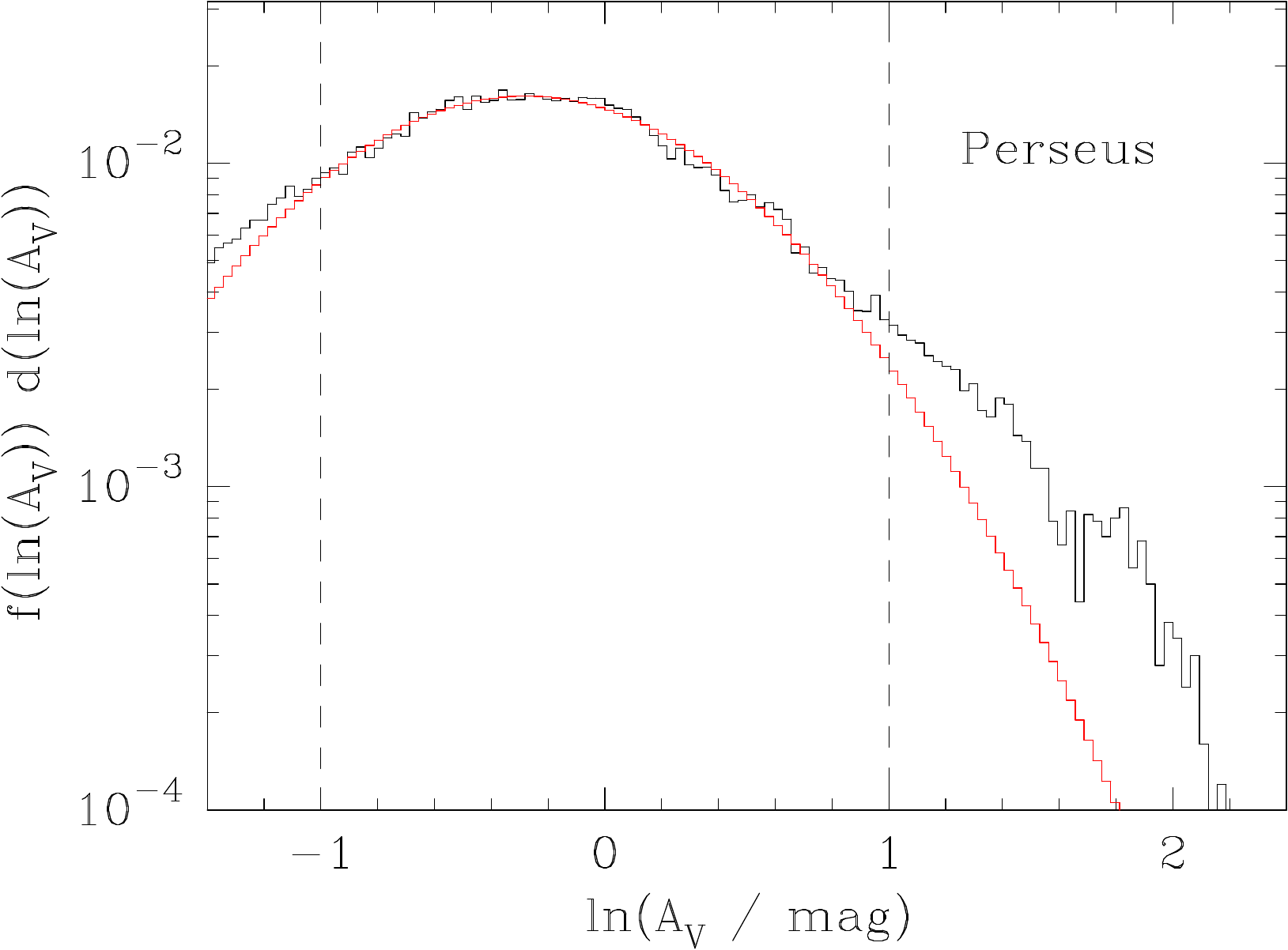}
\includegraphics[width=54mm]{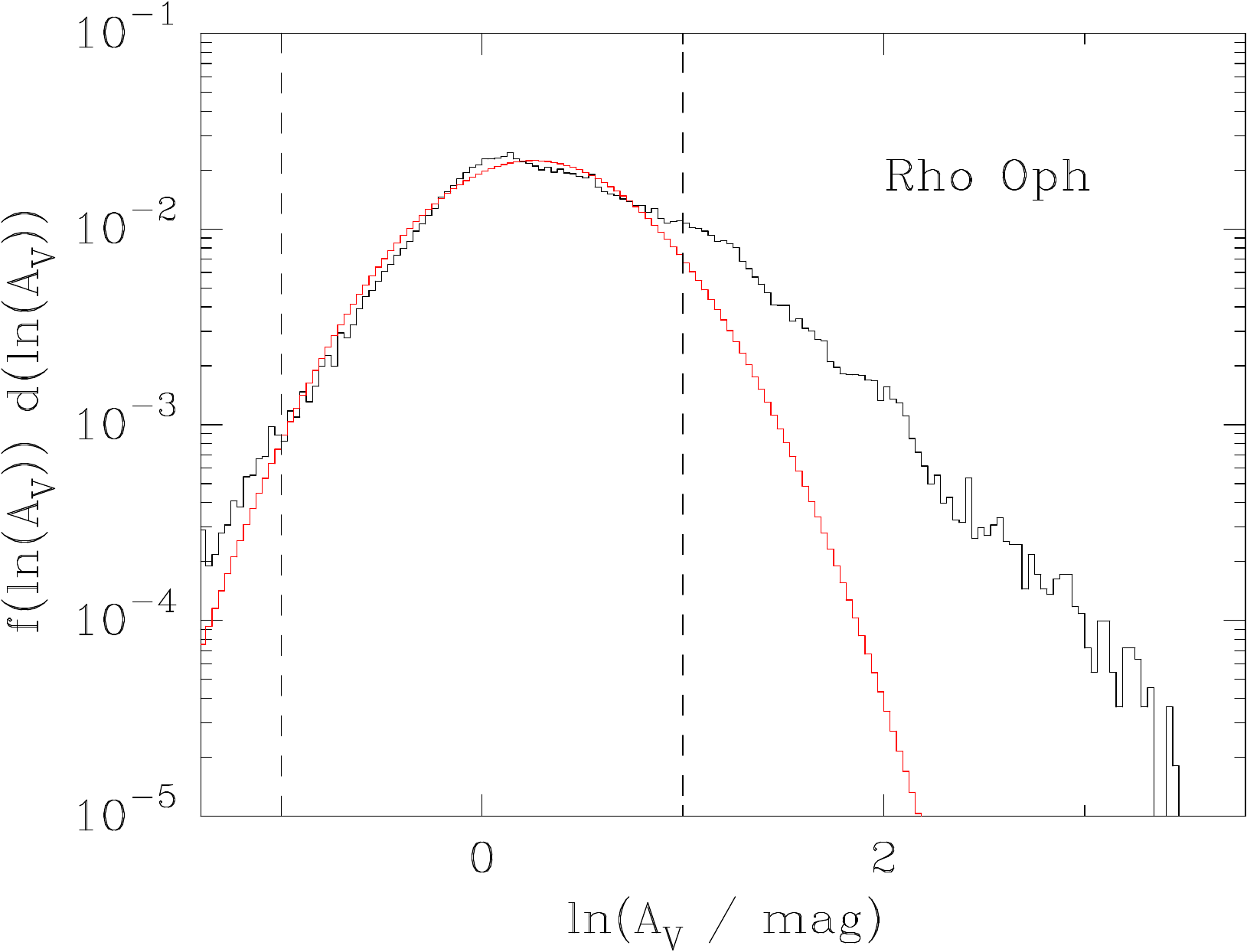}
\caption{Fits (red) to the core of the log space PDFs (black) for the three fields. The fitting range is indicated by the dashed lines.}
\label{fig:pertauophlogplot}
\end{figure*}

\begin{figure*}
\includegraphics[width=54mm]{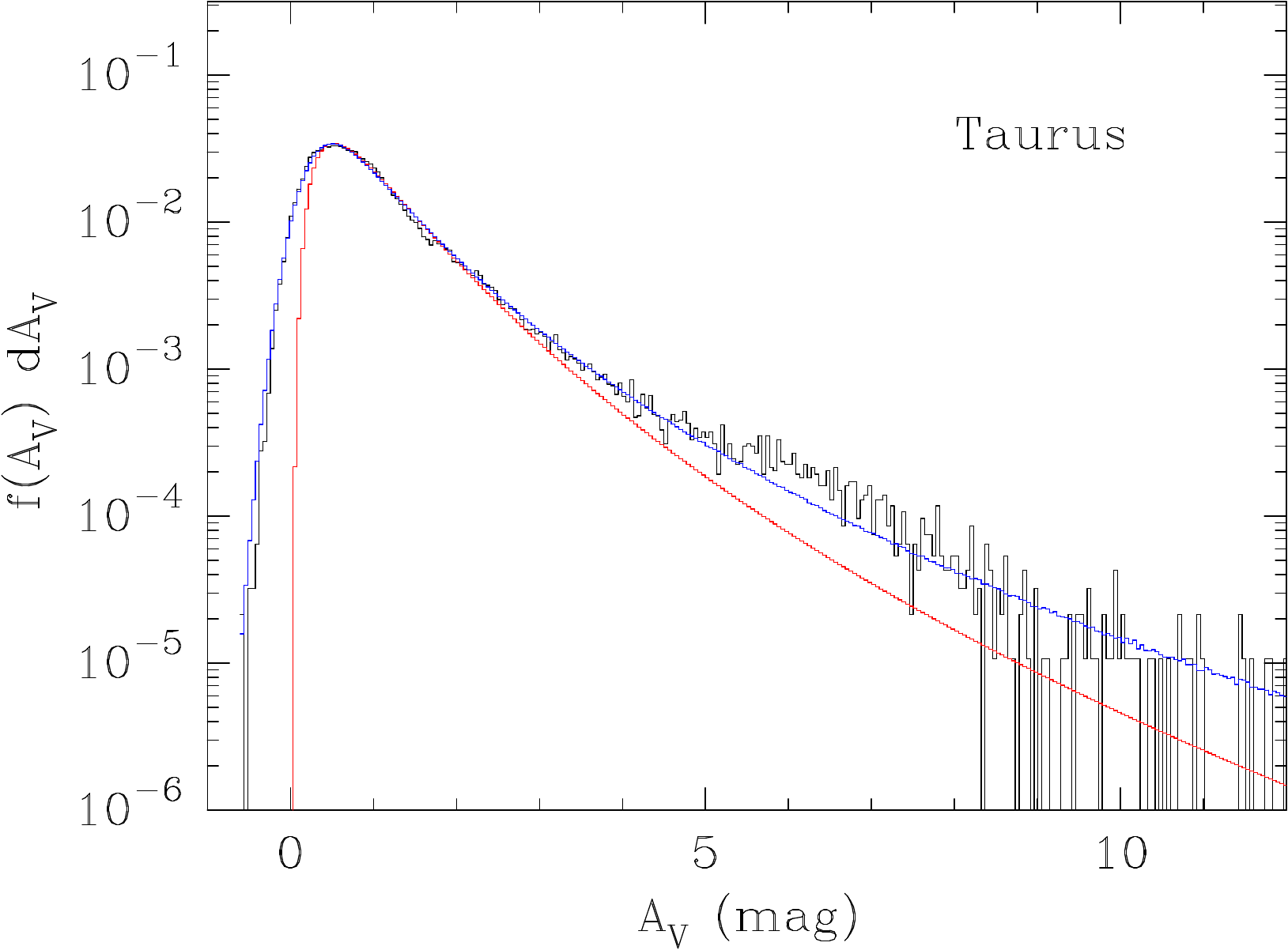}
\includegraphics[width=54mm]{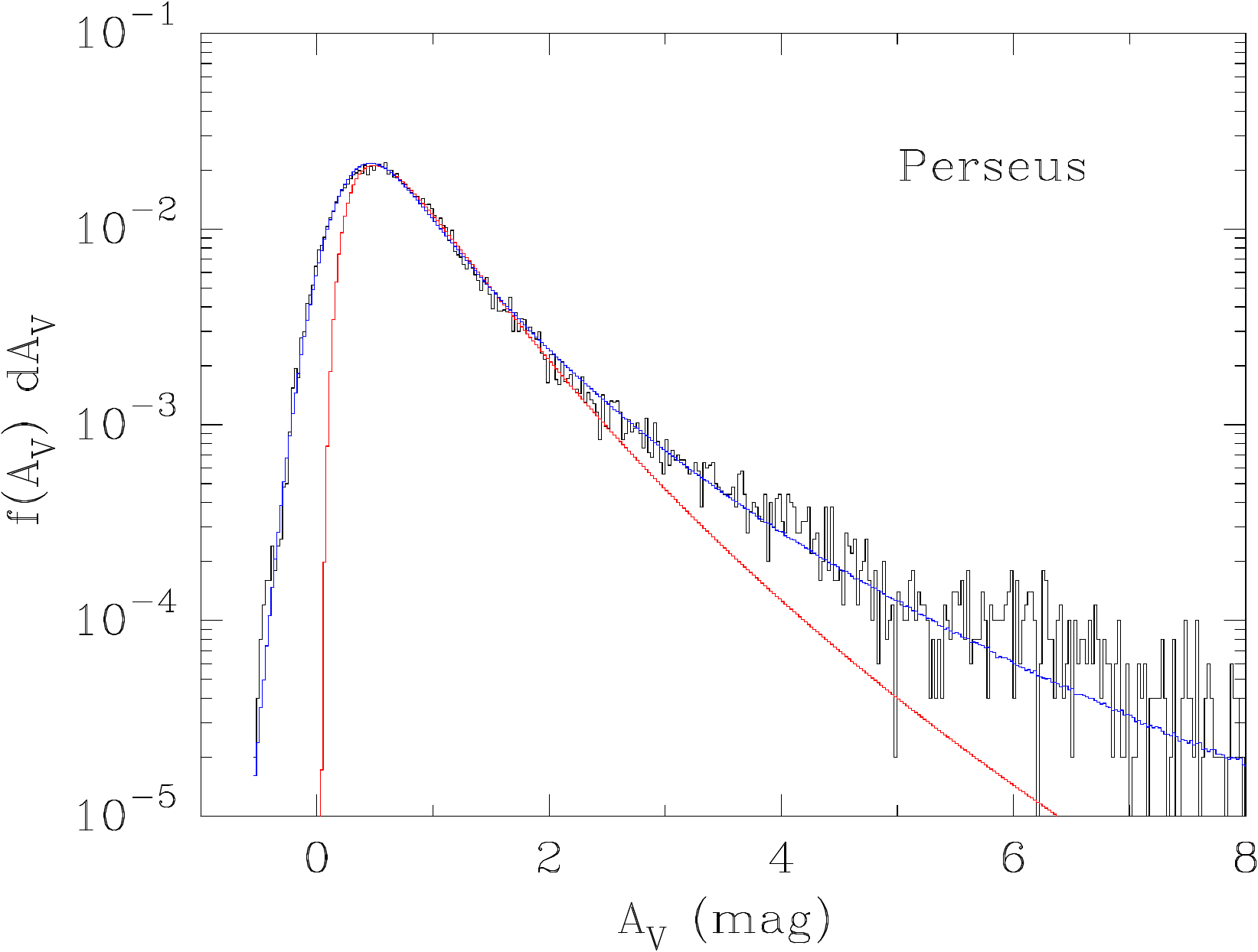}
\includegraphics[width=54mm]{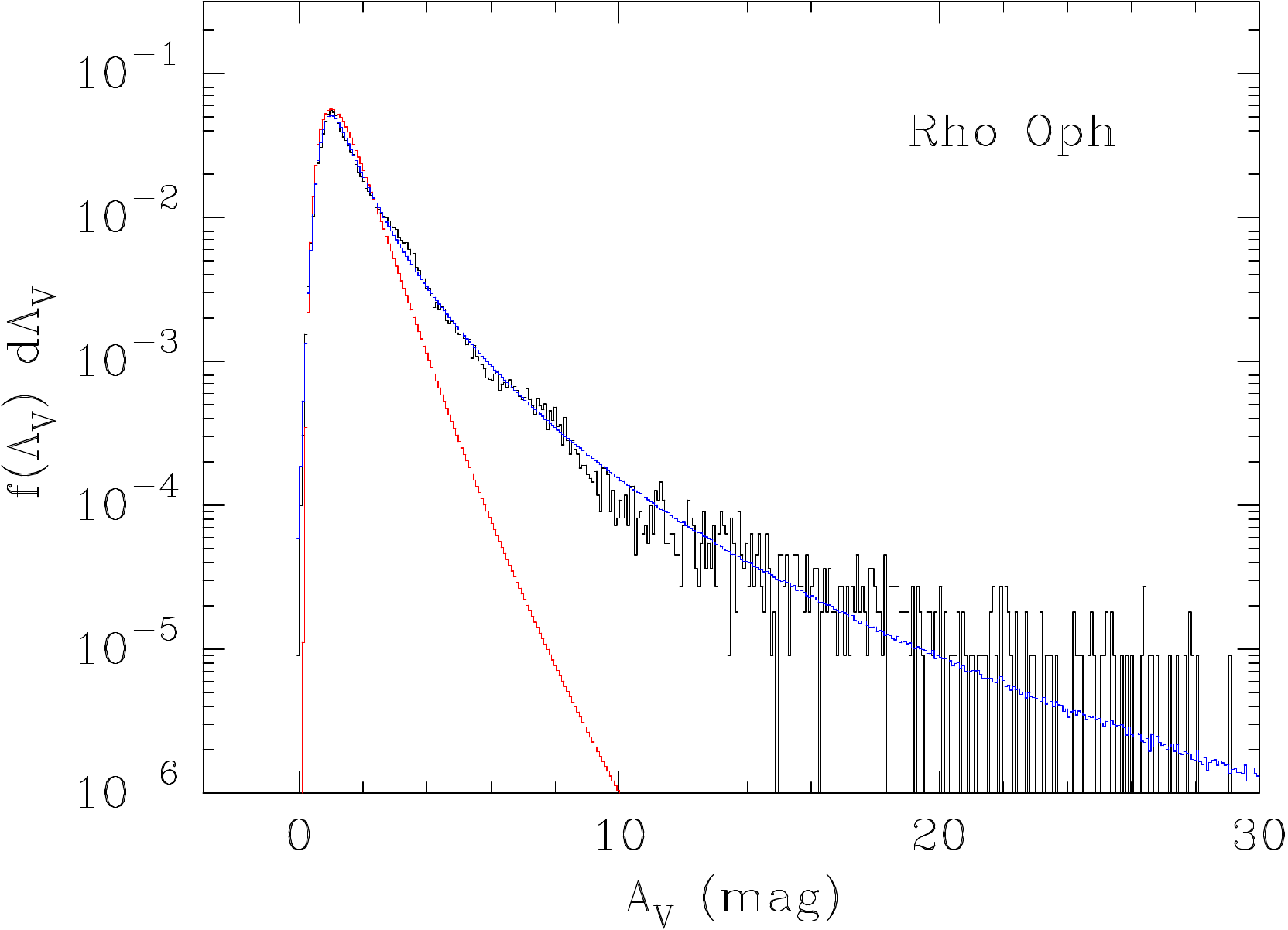}
\caption{Direct fits (blue) to the column density PDFs (black), including noise PDF convolution. The direct space representation of
the fits from Figure~\ref{fig:pertauophlogplot} are shown as red lines. }
\label{fig:pertauophplot}
\end{figure*}

\section{Lognormal PDFs in the presence of noise and background uncertainty}

We will start with a baseline assumption that a column density field can be described by 
a single lognormal function, and evaluate how well this represents the data for a sample of
column density fields determined through extinction mapping. Variations on this scheme will
be presented subsequently.

If the true column density PDF, $f(N)$, is lognormal then $\ln{(N)}$ is normally distributed. Let $\ln{(N)}$
have a mean $\mu$ and variance $\sigma^{2}$. Then the mean and variance of $N$ are, respectively:
\begin{equation}
\mu_{N} = \langle N \rangle = {\mathrm{e}}^{\mu + \sigma^{2}/2} ,
\end{equation}
\begin{equation}
\sigma_{N}^{2} = \langle N^{2} \rangle - \langle N \rangle^{2} = ({\mathrm{e}}^{\sigma^{2}} - 1) {\mathrm{e}}^{2\mu + \sigma^{2}} = ({\mathrm{e}}^{\sigma^{2}} - 1) \mu_{N}^{2} .
\end{equation}
The assumed lognormal $N$-PDF is given by:
\begin{equation}
f(N) = \frac{1}{\sqrt{2{\pi}\sigma^{2}}N} {\mathrm{e}}^{-  \frac{({\mathrm{ln}(N) - \mu)^{2}}}{2\sigma^{2}}} .
\end{equation}

If the column density field is observed in the presence of Gaussian noise with mean zero
and variance $\sigma^{2}_{noise}$, then assuming the noises and column densities are independent,
then the expected observed mean, $\mu_{N,obs}$ and variance, $\sigma^{2}_{N,obs}$, are, respectively:
\begin{equation}
\mu_{N,obs} = \langle N_{obs} \rangle = \mu_{N},
\label{eq:muobs}
\end{equation}
\begin{equation}
\sigma_{N,obs}^{2} = \sigma^{2}_{N} + \sigma^{2}_{noise} .
\label{eq:sigobs}
\end{equation}
From equations~(\ref{eq:muobs}) and (\ref{eq:sigobs}) we can obtain a noise-corrected
estimate of the mean and variance of the noiseless $N$-field if $\sigma^{2}_{noise}$ is known 
(these equations are generally true for independent signal and noise, not just in the case of a lognormal PDF).
The log-space variance and mean are estimated, respectively, by:
\begin{equation}
\sigma^{2} = {\mathrm{ln}} (1 + \sigma_{N}^{2}/\mu_{N}^{2}) ,
\label{eq:sigdef}
\end{equation}
\begin{equation}
\mu = {\mathrm{ln}}(\mu_{N}) - \sigma^{2}/2 ,
\end{equation}
where
\begin{equation}
\mu_{N} = \mu_{N,obs},
\label{eq:muobs2}
\end{equation}
\begin{equation}
\sigma_{N}^{2} = \sigma^{2}_{N,obs} - \sigma^{2}_{noise} .
\label{eq:sigobs2}
\end{equation}

More generally, if each $N$ is uncertain by a noise contribution, then the observed PDF, $f_{obs}$,
is just the true PDF, $f$, convolved with the noise PDF, $f_{noise}$:
\begin{equation}f_{obs} = f \otimes f_{noise} .
\end{equation}

For extinction mapping, the noise variance may not be well-represented by a single value at all extinction levels, but instead
is likely to rise with the extinction level. More elaborate schemes could take this into account, but here we will use
a single value to represent the noise variance, noting that its $\sim$constant value at low extinctions is most important. 

If the observed $N$-field contains a contribution from unrelated background (and/or foreground), 
we can consider the effect of this in simple situations. For a constant foreground/background
contribution, $N_{fb} = \mu_{fb}$, the observed mean is just biased high:
\begin{equation}
\mu_{N,obs} = \mu_{N} + \mu_{fb} .
\end{equation}
The foreground/background may not be exactly constant, but it will likely be significantly less variable than the cloud of interest.
Fitting to the $N$-PDF could include a preliminary subtraction of $\mu_{fb}$ (allowed a free parameter in
the fit). 

\begin{table*}
  \begin{tabular}{lllllll}
    Field & $\sigma_{eff}$ (mag.) & $\mu_{fb}$ (mag.) & $\sigma$ (log core fit) & $\mu$ (log core fit) & $\sigma$ (direct fit) & $\mu$ (direct fit)\\ \hline
    Taurus        & 0.21 & 0.00 & 0.69 & -0.16 & 0.81 & -0.25 \\
    Perseus       & 0.21 & 0.07 & 0.65 & -0.28 & 0.89 & -0.546 \\
    Rho Oph       & 0.21 & 0.54 & 0.49 & +0.26 & 0.97 & -0.17 \\
  \end{tabular}
  \caption{Fitted parameters to the fields. Direct fits  are performed after subtraction of $\mu_{fb}$.}
  \label{tab:table}
\end{table*}

We can absorb some variability in $N_{fb}$ into the noise variance, so that the effective noise variance is:
\begin{equation}
\sigma^{2}_{eff} = \sigma^{2}_{N_{fb}} + \sigma^{2}_{noise} ,
\end{equation}
though this won't be a good representation of gradients in the foreground/background.
For now we just allow variations in $N_{fb}$ to be Gaussian around the mean $\mu_{fb}$. In principle,
$N_{fb}$ could also be lognormally distributed (for example) but for low enough variance around the mean, 
it will appear normal. Below, we allow $\sigma^{2}_{eff}$ to vary around the nominal noise level to
see if it improves the fit.

\section{Fitting to the PDF versus ${\mathrm{\ln}}(N)$}

\begin{figure*}
\includegraphics[width=54mm]{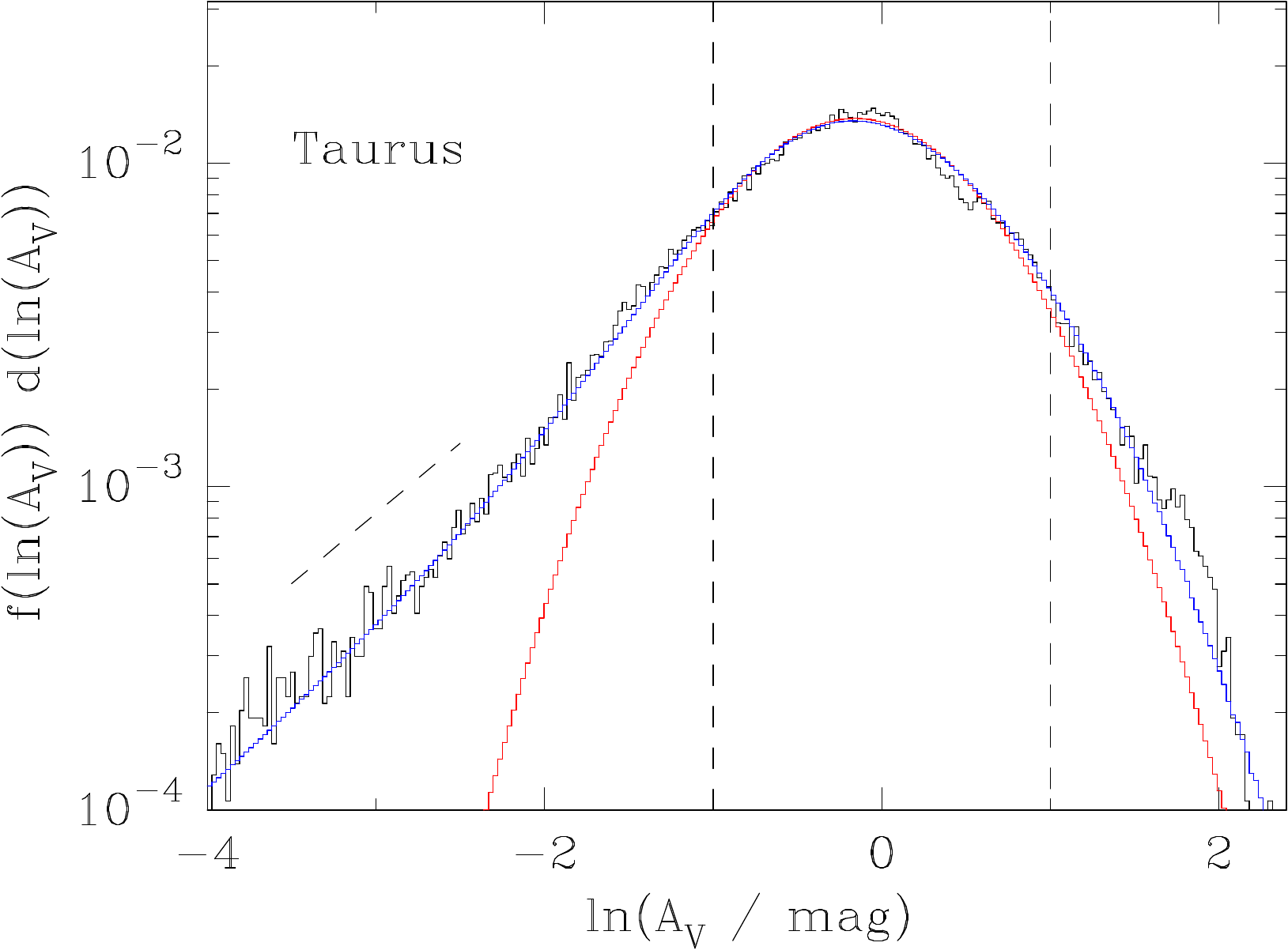}
\includegraphics[width=54mm]{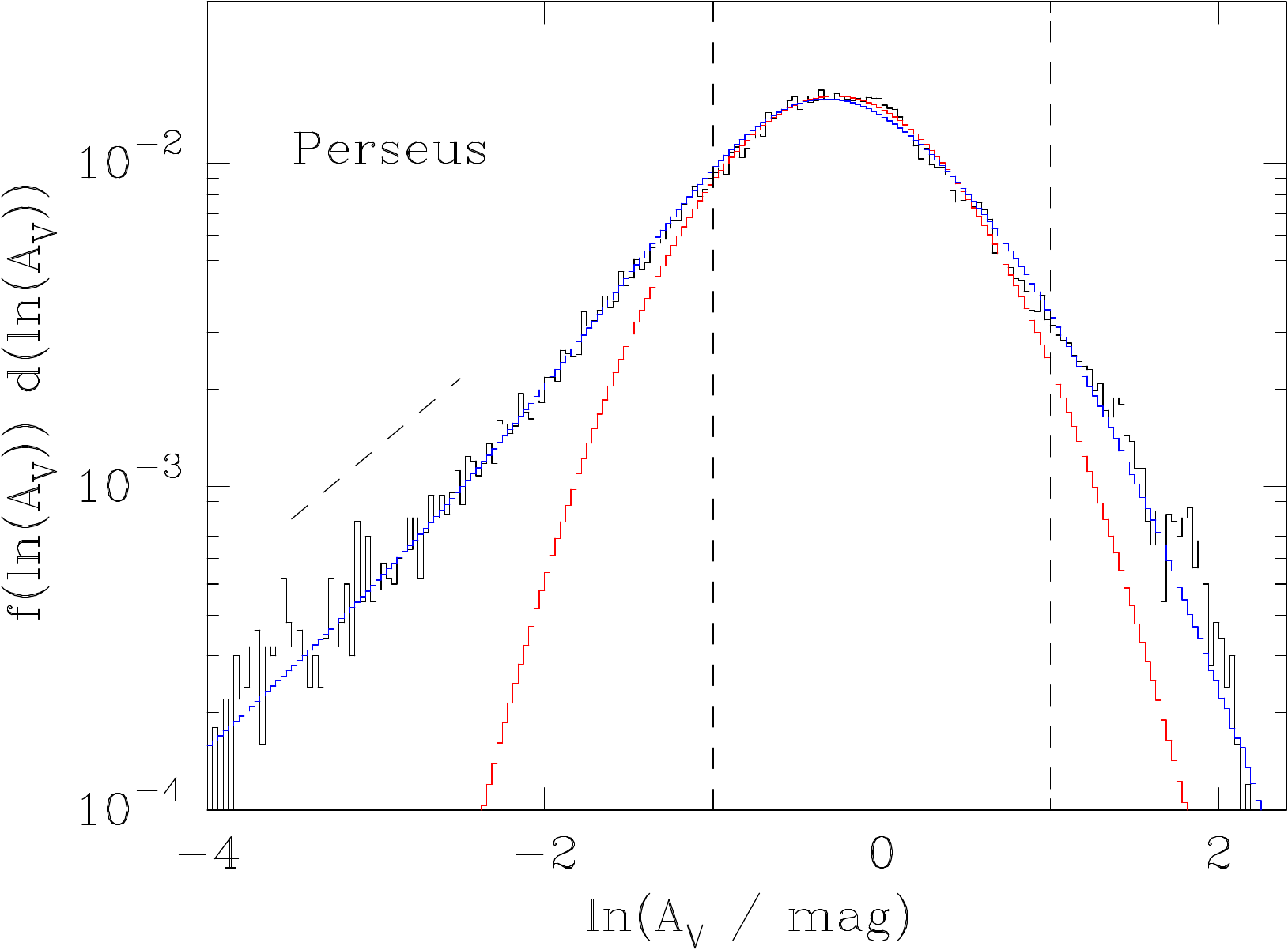}
\includegraphics[width=54mm]{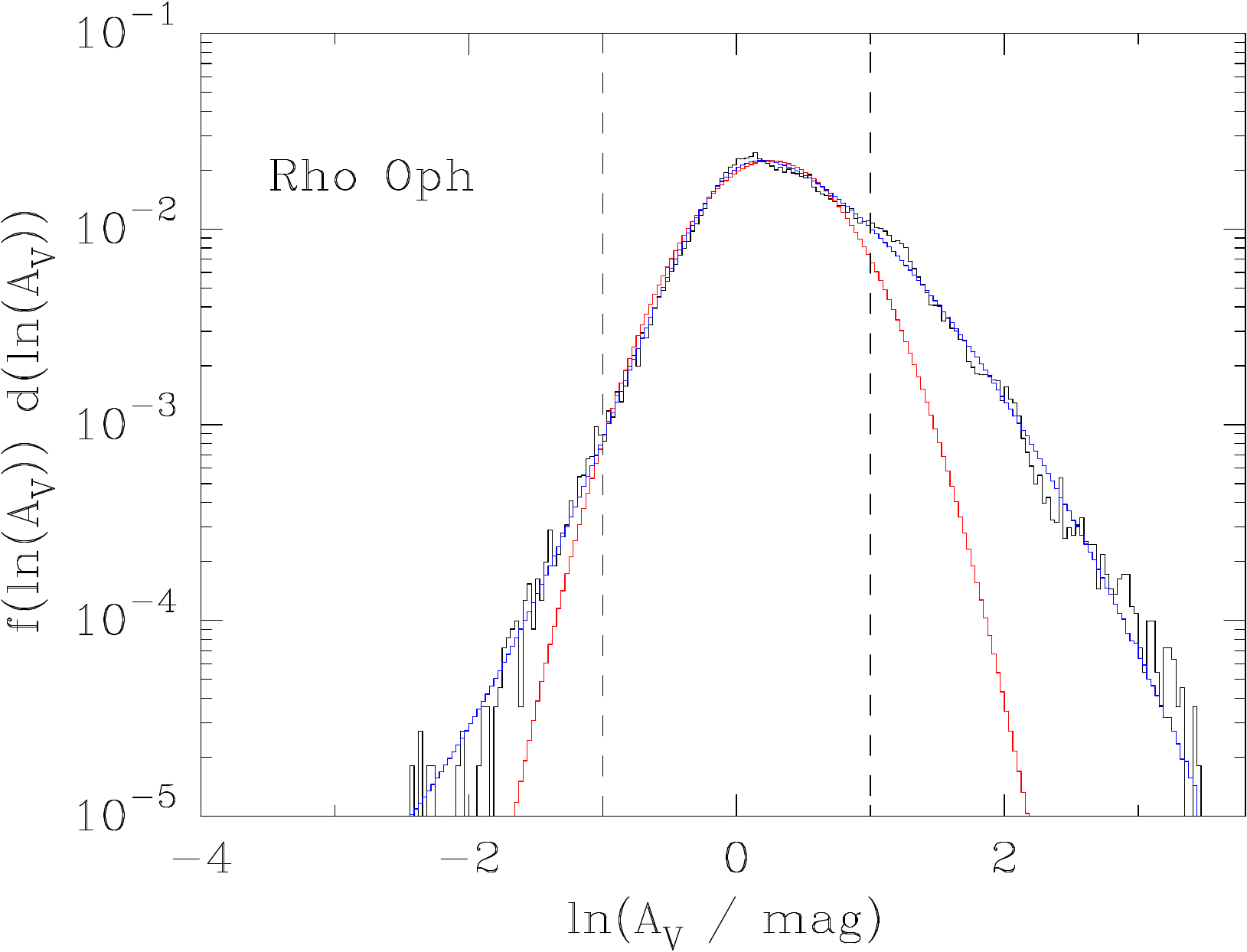}
\caption{Log space representation of the direct lognormal fits (blue) from Figure~\ref{fig:pertauophplot} compared to the log space PDFs (black). The core fits are shown in red. On the Taurus and Perseus plots, a dashed line of slope log$_{10}$(e) is shown (see text).}
\label{fig:pertauophlogplot2}
\end{figure*}

For reference, we perform fits on extinction maps using the widely-used procedure of fitting
cores to the ${\mathrm{\ln}}(N)$-PDFs over a restricted range of ${\mathrm{\ln}}(N)$
where the PDF appears gaussian (i.e. lognormal in $N$).  All fits here are done using A$_{{\mathrm{V}}}$
in place of $N$. We have fitted three fields: Taurus and Perseus, taken from Rowles \& Froebrich (2009)
and Rho Oph, taken from the {\sc complete} database (Ridge et al 2006). The field selection for
these clouds was initially done by eye for Taurus and Perseus -- we will describe in more detail the effects
of field selection in Section~6. For Rho Oph, we just used the full field made available in the
{\sc complete} database, which is closely-concentrated on the molecular cloud.

Figure~\ref{fig:pertauophlogplot} shows the PDFs of ${\mathrm{\ln}}$(A$_{{\mathrm{V}}}$), along with a 
fit to the presumed core region, made between the vertical dashed lines. Excess $\sim$~power-law tails of varying amplitude, 
relative to the fitted lognormals, are seen (c.f. Kainulainen et al 2009). In the fits above, we
have not made a correction for the presence of foreground/background contamination (e.g. Schneider et al 2014, arXiv 1403.2996 submitted).
These are negligibly small for Taurus and Perseus anyway. Neither have we attempted to optimize the fitting range;
this does not affect the major conclusions of the paper.

\section{Direct fitting to $N$ with noise PDF convolution}

Again using A$_{{\mathrm{V}}}$ in place of $N$, we take the observed maps, A$_{{\mathrm{V,obs}}}$, and
varying $\mu_{fb}$ and $\sigma^{2}_{eff}$, calculate:
\begin{equation}
\mu_{A_{{\mathrm{V}}}} = \langle A_{{\mathrm{V,obs}}} \rangle - \mu_{fb} 
\label{eq:muav}
\end{equation}
and
\begin{equation}
\sigma^{2}_{A_{{\mathrm{V}}}} = \sigma^{2}_{A_{{\mathrm{V,obs}}}} - \sigma^{2}_{eff} . 
\label{eq:sigav}
\end{equation}
We then create a lognormal PDF and convolve it with a noise PDF of variance $\sigma^{2}_{eff}$,
looking for the values of $\mu_{fb}$ and $\sigma^{2}_{eff}$ that provide the best fit to the
observed PDF. 
Note that once $\mu_{A_{{\mathrm{V}}}}$ and $\sigma^{2}_{A_{{\mathrm{V}}}}$ are calculated, no
further optimisation of the fit by varying these estimates is performed, i.e. 
(for these initial fits) equations~(\ref{eq:muav}--\ref{eq:sigav}) hold exactly. 

We find that $\sigma_{eff} = 0.21$~mag. provides the best fit for all fields: this is comparable
to the noise estimates given in the data sources, and comparable to direct estimates from the fields
themselves in ``uninteresting'' regions. Note that $\sigma^{2}_{eff}$ does not include calibration
uncertainties, but is just an estimate of the noise variation in the maps.

\begin{figure*}
\includegraphics[width=174mm]{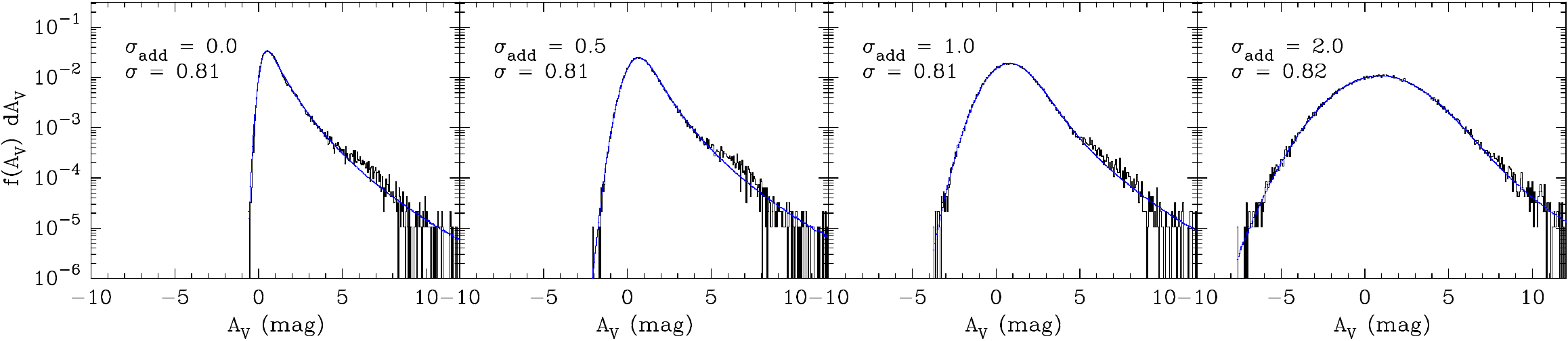}
\caption{Direct space fits (blue lines) to the Taurus PDF with added noise (black lines). The best fit lognormal $\sigma$ is quoted in each panel.}
\label{fig:taunoisefits}
\end{figure*}

\begin{figure*}
\includegraphics[width=174mm]{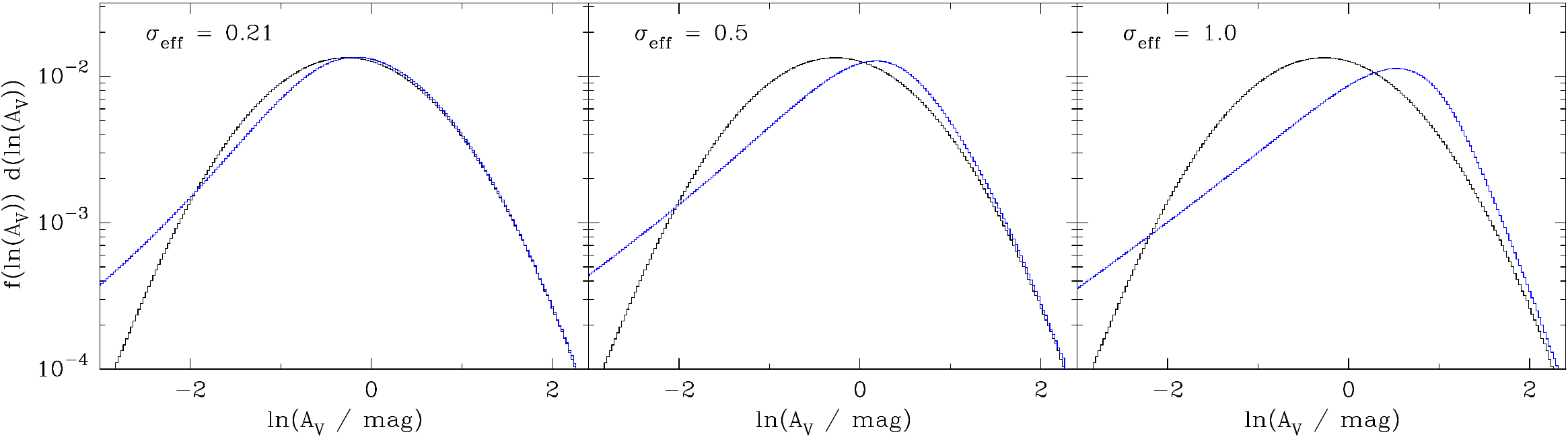}
\caption{Log space lognormal PDFs with noise added (blue lines) compared to the noise-free fitted PDF of Taurus (black lines).}
\label{fig:noisepdfdata}
\end{figure*}

\begin{figure*}
\includegraphics[width=174mm]{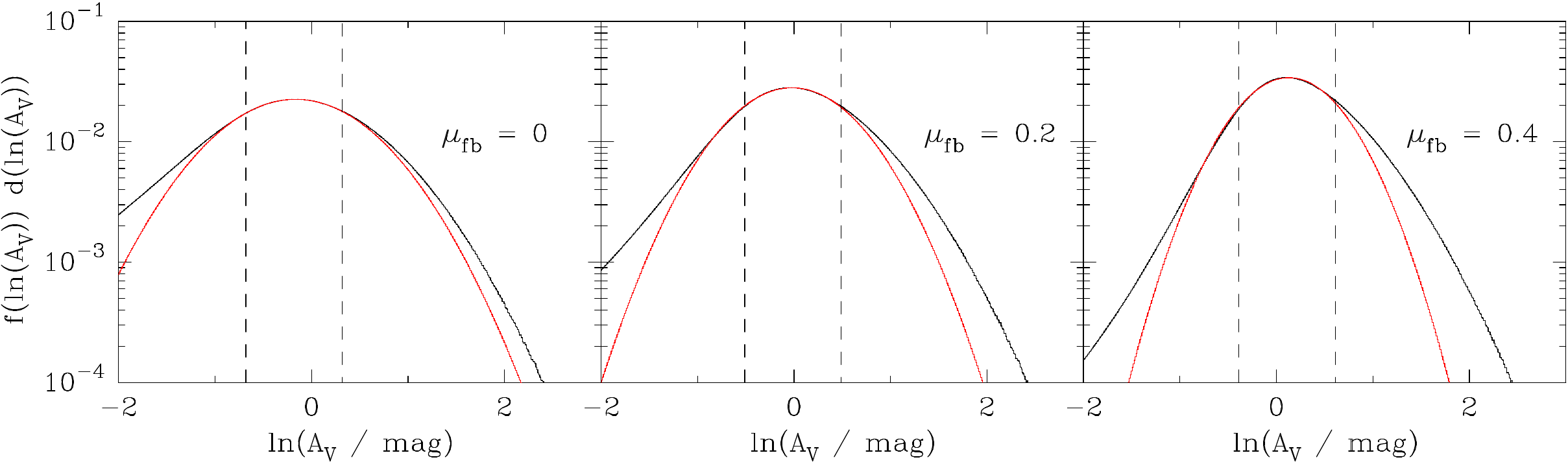}
\caption{Effect of an unaccounted-for foreground/background level ($\mu_{fb}$). The noise-convolved lognormal fit to Taurus, subjected to
different $\mu_{fb}$ values is shown as black lines. Fits to the core are shown as red lines. }
\label{fig:logdistort}
\end{figure*}

Little or no foreground/background correction is necessary for the Rowles \& Froebrich maps {as reference fields
are used in the construction of these maps and they lie off the Galactic Plane near the Galactic anti-centre}, 
but we find that the {\sc complete} Rho Oph map is best fitted with a constant foreground/background level of 
$\mu_{fb} = 0.54$~mag. (which is small compared to the A$_{{\mathrm{V,obs}}}$ in the main Rho Oph cloud). 
No reference-level subtraction is reported in Ridge et al (2006) but their extinction
map is calibrated to IRAS-derived 100~$\mu$m optical depth and therefore will necessarily include contributions
from the full dust columns along the line-of-sight. Our estimated foregound/background of 0.54~magnitudes implies
an E(B--V)~$\approx$~0.17 (R = 3.1). This is consistent with the Schlegel, Finkbeiner \& Davis (SFD; 1998) E(B--V)
map on the periphery of the Rho Oph field which has E(B--V) values in relatively unstructured regions
towards higher/lower Galactic latitude of $\sim$~0.1 and 0.25 respectively. The foreground/background across 
Rho Oph is therefore not constant but contributes an additional dispersion above the Ridge et al (2006) 
nominal dispersion of 0.16 magnitudes (a dispersion of 0.21 magnitudes is measured directly in the field). 
If normally-distributed, the foreground/background fluctuations should have a
dispersion in E(B--V) of $\sim$~0.044, which is comparable to that measured directly in the SFD map
over scales comparable to the Rho Oph cloud (a dispersion of $\sim$~0.055 is measured directly; accounting for 
a 16\% internal calibration dispersion, this predicts a natural dispersion of 0.048 for a mean E(B--V) level of 0.17).

The resulting PDF estimates, after convolution with the noise PDF are shown in Figure~\ref{fig:pertauophplot}.
We also convert the gaussian fits to the ${\mathrm{\ln}}$(A$_{{\mathrm{V}}}$) core into
the direct A$_{{\mathrm{V}}}$ representation and show these for comparison.
All fitted parameters are listed in Table~\ref{tab:table}, along with the log core fits of the previous Section.

We convert the noise-convolved PDFs of A$_{{\mathrm{V}}}$ into the log representation for
inspection. These are shown in Figure~\ref{fig:pertauophlogplot2}, and emphasize the excellence of the
fit, both in the high A$_{{\mathrm{V}}}$ tail and the noise-dominated low A$_{{\mathrm{V}}}$ tail. (We will
say more quantitatively about how good the fits are in the following Sections.) The appearance of a
lognormal PDF with noise and/or offset is no longer symmetric about its maximum in the log representation; i.e. it is 
no longer a pure parabola.
Counterintuitively, the width of the log space PDF near its peak is in fact {\it narrowed} by the noise convolution 
(see next Section), and the core fits to this region in the log representation are not reliable without 
taking into account the noise smearing. 

In the log representation, the asymptotic slope of the negative PDF tail can be estimated, as long as the
amplitude of the PDF is significant near A$_{{\mathrm{V}}} \approx 0$ due to noise.
Note that, as $\ln$(A$_{{\mathrm{V}}}$) becomes more negative, equal intervals in $\ln$(A$_{{\mathrm{V}}}$)
correspond to progressively smaller intervals in A$_{{\mathrm{V}}}$ as it approaches zero. In the presence of noise, if we
approximate the observed PDF as $f$(A$_{{\mathrm{V}}}$)$\sim$constant over these small intervals, this predicts:
\begin{equation}
f(\ln(A_{{\mathrm{V}}})) = f(A_{{\mathrm{V}}}) \frac{{\mathrm{d}}A_{{\mathrm{V}}}}{{\mathrm{d}}\ln(A_{{\mathrm{V}}})}    \propto A_{{\mathrm{V}}} \propto {\mathrm{exp}}(\ln(A_{{\mathrm{V}}})) 
\end{equation}
in the negative log-space tail. In a plot of log$_{10}$(f($\ln$(A$_{{\mathrm{V}}}$))) versus $\ln$(A$_{{\mathrm{V}}}$)
(Figure~\ref{fig:pertauophlogplot2}) the aysmoptotic slope is predicted to be log$_{10}$(e)~$\approx$~0.43. A dashed
line with slope log$_{10}$(e) is shown in the Taurus and Perseus panels, and agrees well with the observed slopes. 
For Rho Oph, the presence of a significant foreground has resulted in a low observed PDF amplitude near A$_{{\mathrm{V}}} \approx 0$
so that the above approximation is inapplicable.  A low PDF amplitude near A$_{{\mathrm{V}}} \approx 0$ may also 
occur if the PDF is sampled only towards high column density regions (see e.g. Section~6) so some care in
interpreting the asymptote is needed.
 
The moment-based fitting method has been also used recently by Butler, Tan, \& Kainulainen (2014) and 
Kainulainen \& Tan (2013). Butler et al (2014) do not find evidence for a power-law tail in their data,
while Kainulainen \& Tan (2013) show that the PDFs of their analyzed Infrared Dark Clouds, as well as 
Rho~Oph, are consistent with single (broad) lognormal PDFs. In fact, in their re-analysis of the Kainulainen et al (2009) 
extinction data, Kainulainen \& Tan (2013) present noise-corrected values for $\sigma_{N}/\mu_{N}$, which lead to 
estimates for $\sigma$ of 0.77, 0.83, and 0.89 in Taurus, Perseus and Rho~Oph respectively; these are  
similar to our values (0.81, 0.89, and 0.97) reported in Table~\ref{tab:table}, though a little lower 
despite higher resolution. This may be due to a 
slightly higher foreground/background contribution in their data. The inferred $\sigma$ from
Kainulainen \& Tan (2013) from moment analysis exceed the core-fitted values in Kainulainen et al (2009), 
by factors $\sim$2. In general however, there is no definitive value for $\sigma$ that can be extracted
from extinction maps, as we demonstrate below (Section~8).

\section{Uncertainties and Excesses}

\subsection{Effects of noise}

The method above is stable against noise. Negative column densities 
caused by noise pose no problem: they need not be excluded from the fit, and they contribute to
constraining the low end of the PDF. 
In the log representation, very low and negative noise values are obviously problematic.
Schneider et al (2014) have outlined a very careful scheme for dealing with noise
and foreground/background contamination in log space, which relies on empirical corrections
based on the shape and amplitude of the low A$_{{\mathrm{V}}}$ tail in log space. From
the direct space perspective, this is accounted for in the fitting directly, using all the data.

If the noise is Gaussian, then the leading uncertainties due to noise are the uncertainties
in the noise's zero level ($\pm \sigma_{eff}/\sqrt{N_{pix}}$ for $N_{pix}$ pixels in the map) and
the noise variance ($\pm \sqrt{2}\sigma_{eff}^{2}/\sqrt{N_{pix}-1}$). For a large enough map, these uncertainties
are negligible.  For large $N_{pix}$ and Gaussian noise, the uncertainty on the lognormal $\sigma$ is
$\pm (\sigma_{eff}/N_{pix}) ( 1/ \mu_{A_{{\mathrm{V}}}}^{2} + 
2 / \sigma_{A_{{\mathrm{V}}}}^{2})^{1/2}$. This is very small ($\sim$~10$^{-3}$) 
for the maps above. 

To show the effect of noise, we added Gaussian noise of standard deviation 
$\sigma_{add}$ = 0.5, 1.0, and 2.0~magnitudes to the Taurus map.
Figure~\ref{fig:taunoisefits} shows
fits obtained in direct space under the assumption that $\sigma_{eff}$ is known. The uncertainty
in the lognormal $\sigma$ rises to $\sim \pm 0.01$ for $\sigma_{add} \sim 2$, in line with
the above estimate. Note that the original $\sigma_{eff}$ in the field contributes
negligibly to the total noise variance in these tests, so imprecise knowledge of its value
is irrelevant. Of course, as the noise variance increases, the detailed structure of the PDF, 
including any deviations from lognormality, is blurred out, and all we are able to
extract are the parameters of the best-fitting lognormal. The effect of noise on the PDF
in direct space is easily discerned: first, the PDF is broadened by the noise
convolution, and second, the PDF peak is shifted upwards in A$_{{\mathrm{V}}}$ from its original peak 
(always below the lognormal's mean) towards the lognormal's mean. 

\begin{figure*}
\includegraphics[width=84mm]{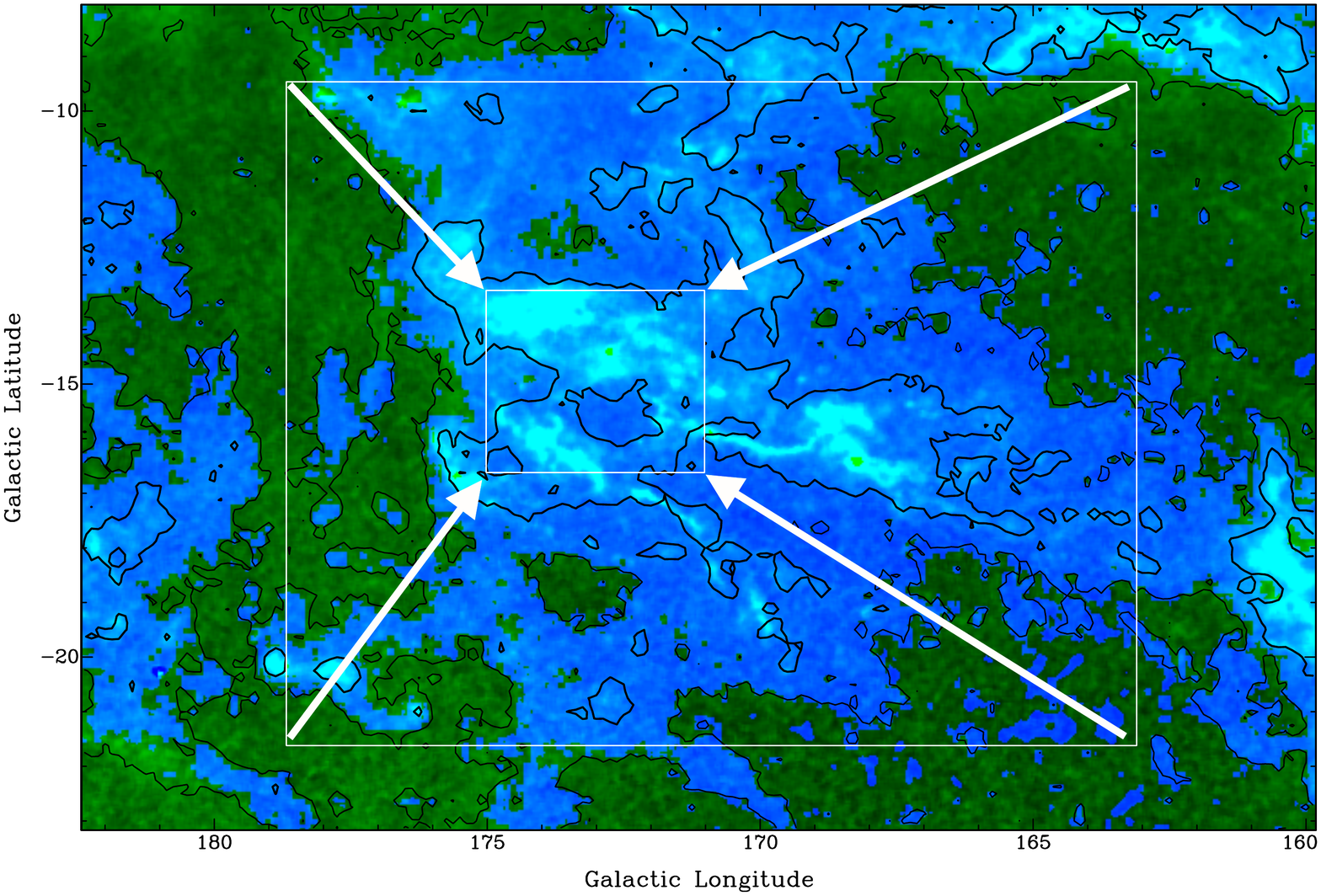}
\includegraphics[width=84mm]{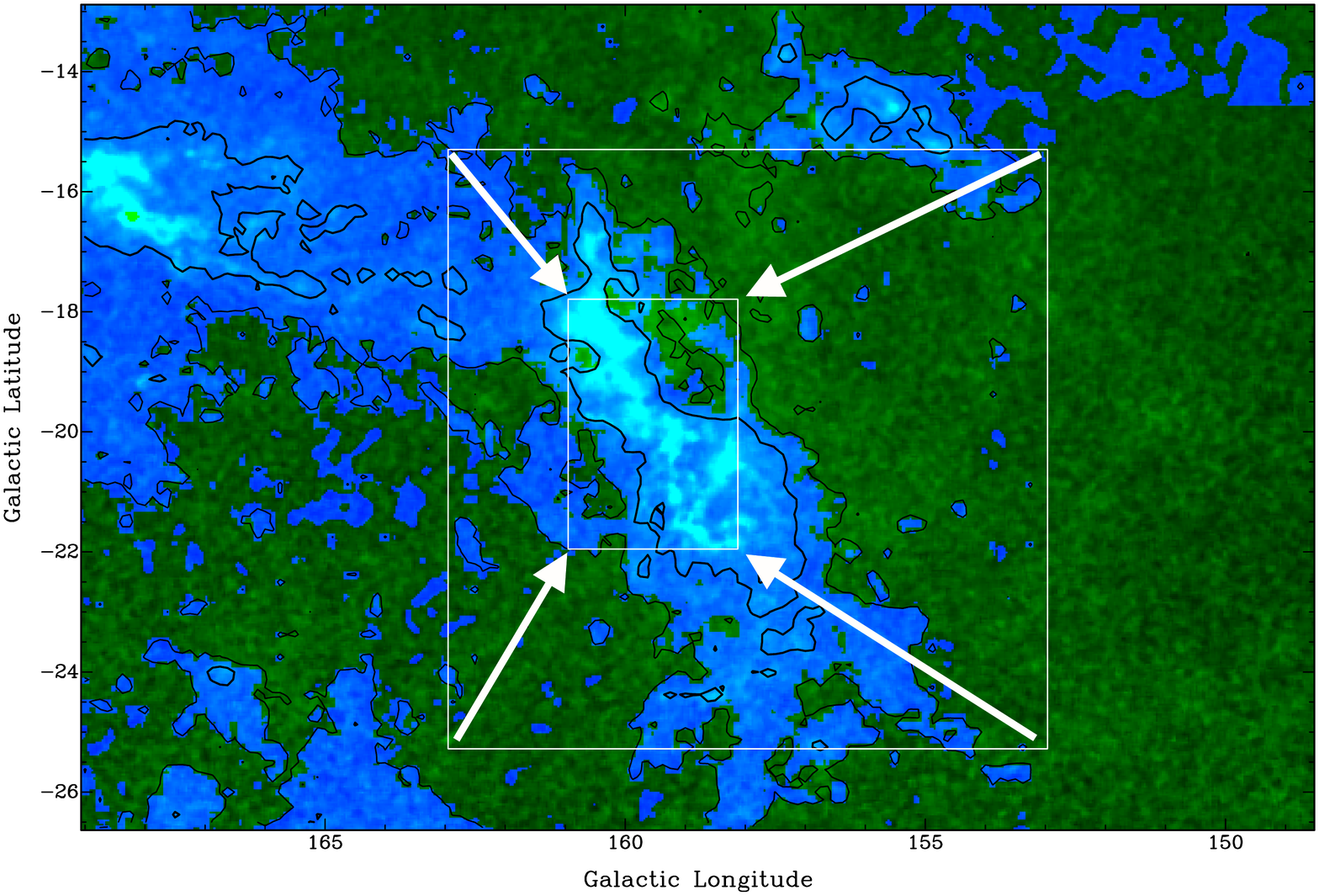}
\includegraphics[width=84mm]{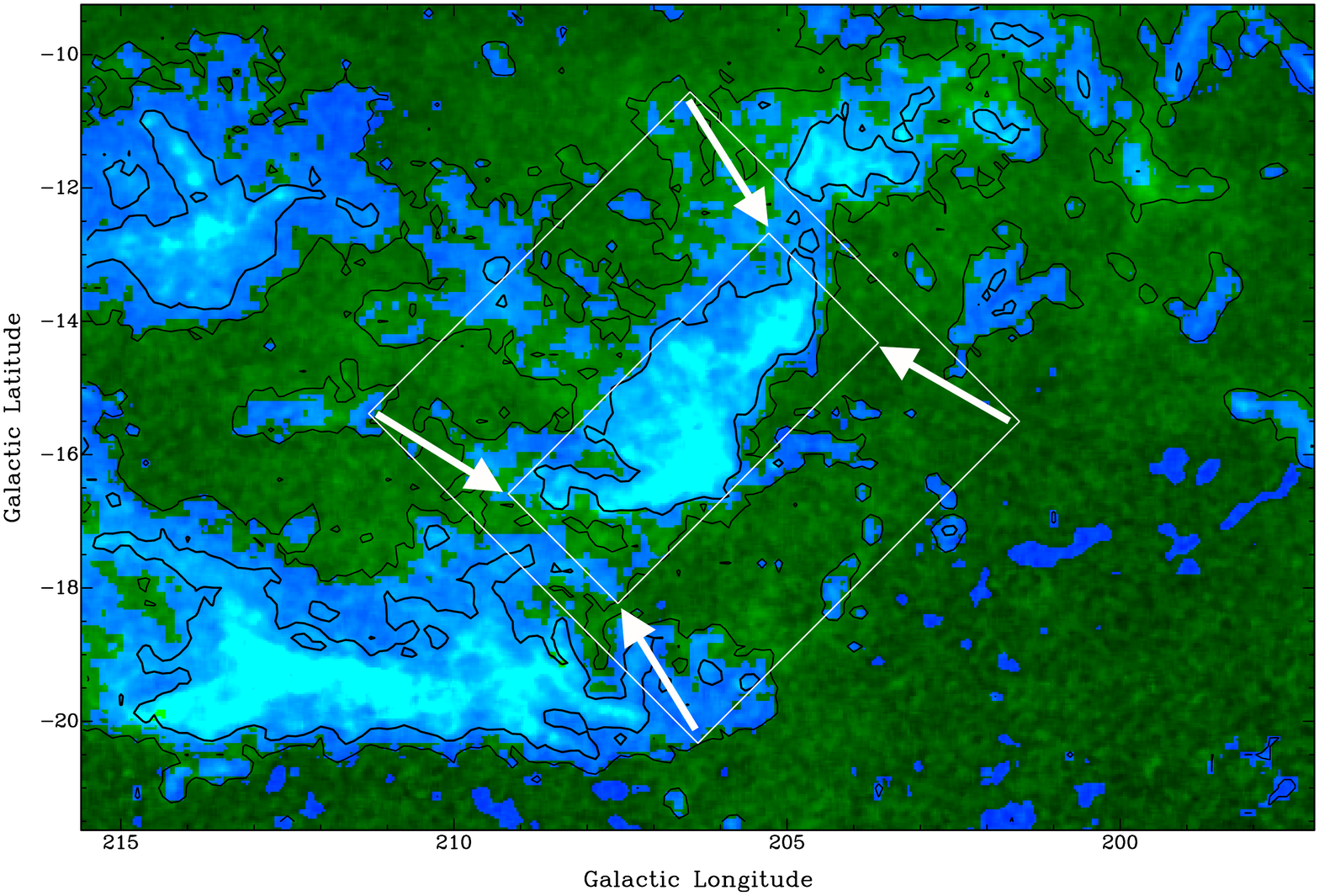}
\includegraphics[width=84mm]{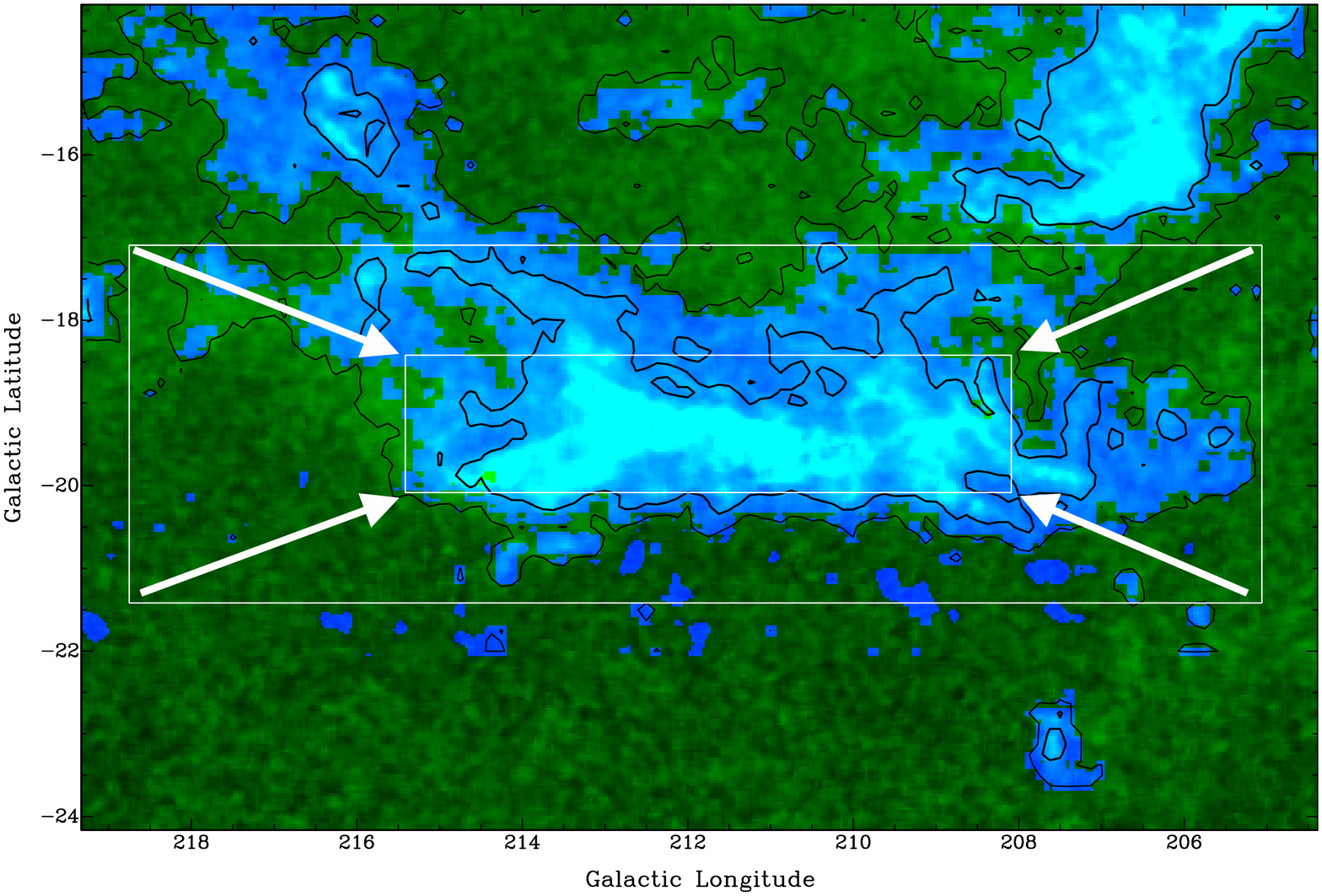}
\caption{Wide and narrow field selections for Taurus, Perseus, Orion B, and Orion A (clockwise from top left).
The arrows denote the direction in which the fields are restricted from wide to narrow. Black contours are at
W$_{{\mathrm{CO}}}$~=~1~K~km~s$^{-1}$ and 10~K~km~s$^{-1}$. Pixels determined to be X-dominated (see text) are coloured blue,
with ``diffuse'' pixels coloured green. The X-dominated boundary roughly coincides with the W$_{{\mathrm{CO}}}$ = 1~K~km~s$^{-1}$ contour.
The ``cold'' molecular part of the extinction is taken to lie within the W$_{{\mathrm{CO}}}$ = 10~K~km~s$^{-1}$ contour.}
\label{fig:fieldselect}
\end{figure*}

The broadening and shifting of the PDF in direct space have counterintuitive effects in the
log representation. The shift of the direct space PDF in fact causes the log space PDF to
become {\it narrower} (lower apparent $\sigma$) near its peak, which is also shifted. To show this,
we plot in Figure~\ref{fig:noisepdfdata} the PDF fit to Taurus {\it before} noise convolution (i.e.
an exact lognormal with parameters given in Table~\ref{tab:table}), compared
to the PDF subjected to noise convolution, using $\sigma_{eff}$~=~0.21, 0.5, and 1.0~magnitudes. 
At $\sigma_{eff}$~=~0.21 (the estimated noise level in the data), the log space PDF at 
$\ln{{\mathrm{A}}_{{\mathrm{V}}}}$ just below the peak is {\it suppressed}, while the higher 
$\ln{{\mathrm{A}}_{{\mathrm{V}}}}$ levels are affected much less. Thus the PDF is narrower around
its peak. As $\sigma_{eff}$ increases, the shift and distortion become greater, and the
PDF narrowing around the peak is more emphasized (see also Fig.~5 of Schneider et al 2014).
It is always possible to fit a parabola
close to a smooth local peak, so, without taking into account the noise at the
outset, the PDF fits to the core of the PDF will always be narrower (lower apparent $\sigma$)
than the true PDF, and spurious tails will be created or any existing tails will be
amplified.

\subsection{Effects of foregound/background contributions}

The presence of an unaccounted for foreground/background contribution can have significant
influence on the  PDF in the log representation (Schneider et al 2014). In Figure~\ref{fig:logdistort} we show the
(direct space) fitted PDF to Taurus, including the noise PDF convolution, subjected
to small offsets of $\mu_{fb}$ = 0.0, 0.2, and 0.4~magnitudes, and transferred to the log representation.
As noted above, fitting to the core, even for $\mu_{fb} = 0.0$~mag., leads to spurious excess in the tail,
but this effect is amplified in the presence of $\mu_{fb} > 0$. While the direct space effect is just a 
linear shift along the A$_{{\mathrm{V}}}$ axis, the log space effect is to shift and {\it distort} the PDF (i.e.
similar to the effect of noise shifting the PDF peak).
Fits made to the core as $\mu_{fb}$ increases systematically produce greater excess in the tail. 
It is therefore important to account for possible foreground/background contamination, especially
for more distant clouds, as the contaminating extinction will grow with distance.
Extinction maps made using reference fields will alleviate this problem, but column density
maps made from far infrared emission will be sensitive to the entire dust column and therefore be more
susceptible to this bias. Schneider et al (2014), using {\it Herschel} imaging, indeed find foreground/background
contributions of between 0.8 and 3.0 magnitudes, albeit for more distant clouds than analyzed here (0.45 - 7~kpc).

\section{Field selection}

If the signal-to-noise level is high, the leading uncertainties on the field's mean and 
variance estimates come from the {\it field selection} itself: i.e. exactly how the 
field is chosen (this can be somewhat arbitrary) effectively sets the mean and variance. The
other important source of uncertainty (once the mean and variance are estimated) is
the applicability of the lognormal model. Above, we have used this to predict
$\mu_{A_{{\mathrm{V}}}}$ and $\sigma^{2}_{A{{\mathrm{V}}}}$, but these will differ
from the predictions if the field is not lognormal. As the lognormal model appears
to be well-motivated, this is a good first approximation, following which any
deviations can be accounted for if necessary (e.g. the addition of a power-law component, including
any modifications to the $\mu_{A_{{\mathrm{V}}}}$ and $\sigma^{2}_{A{{\mathrm{V}}}}$ predictions).

We note that while a lognormal can appear linear in the log representation over
limited ranges, it cannot account for positive log-PDF curvature, or account for true power-law
behaviour indefinitely. 
The PDFs shown in the previous Section in fact only displayed minor ``core+tail'' structure; i.e. the PDFs
do not show a very distinct kink at any A$_{{\mathrm{V}}}$ level.
There are some observed PDFs (e.g. Kainulainen et al 2009; Alves de Oliveira 2014; Schneider et al 2014) which display a more 
obvious core+tail behaviour, involving a more distinct high A$_{{\mathrm{V}}}$ departure from the low A$_{{\mathrm{V}}}$ behaviour.
The high A$_{{\mathrm{V}}}$ departure from the low A$_{{\mathrm{V}}}$ lognormal core is not always an obvious power-law however,
but often has the appearance of the high-A$_{{\mathrm{V}}}$ end of a broad lognormal function. 

\subsection{Effects of field selection}

An extinction map does not distinguish between molecular and atomic and gas, while both will in general
be present in the map. Perhaps a more sensible distinction to make is between cold ($\sim$isothermal) gas and warm(er) gas. 
This latter component we will refer to as ``diffuse'' and note that it could include both atomic and warm molecular material (e.g. CO-dark
molecular gas; e.g. Langer et al 2014). 

The low A$_{{\mathrm{V}}}$ structure of observed PDFs will be controlled by the amount of diffuse material
incorporated into the field selection. For different field selections, the amount of diffuse material included will
vary, and therefore the amplitude of the ``core'' will vary. To demonstrate this, we have taken a more general
approach to field selection, as shown in Figure~\ref{fig:fieldselect}. Here we show the Taurus and Perseus molecular clouds
circumscribed by a ``wide'' field selection and a ``narrow'' field selection. We have also included Orion~A and B from
Rowles \& Froebrich (2009) in this analysis, since these clouds display a more distinct ``core+tail'' PDF structure in 
Kainulainen et al 2009. 

Between the wide and narrow field selections, we impose a series of intermediate selections, with the direction
of contraction of the field selection schematically represented by the arrows in Figure~\ref{fig:fieldselect}.  For each selection,
the mean and noise-corrected variance are calculated, a matching lognormal PDF is created and convolved with the noise PDF to
produce a test PDF, $f_{test}$, which is then compared to the observed PDF, $f_{obs}$. We have opted for a simple
rectangular field selection, to avoid the obvious problems that would arise in employing a cut in A$_{{\mathrm{V}}}$ to
define the cloud.

\begin{figure}
\includegraphics[width=84mm]{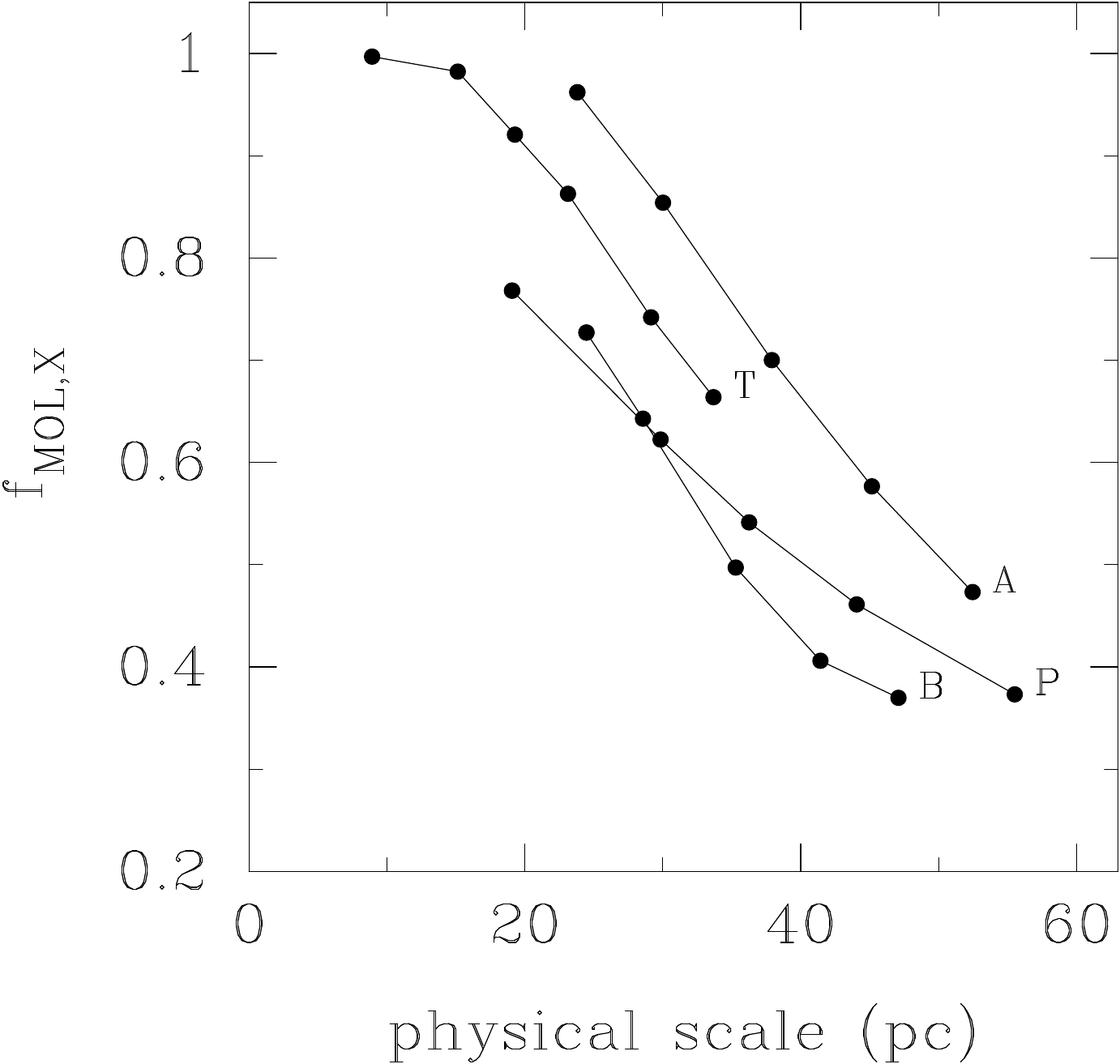}
\caption{X-dominated fraction, $f_{MOL,X}$ versus spatial scale for the field selections in Taurus (T), Perseus (P), Orion A (A), and Orion B (B).}
\label{fig:fmol}
\end{figure}

\begin{figure}
\includegraphics[width=84mm]{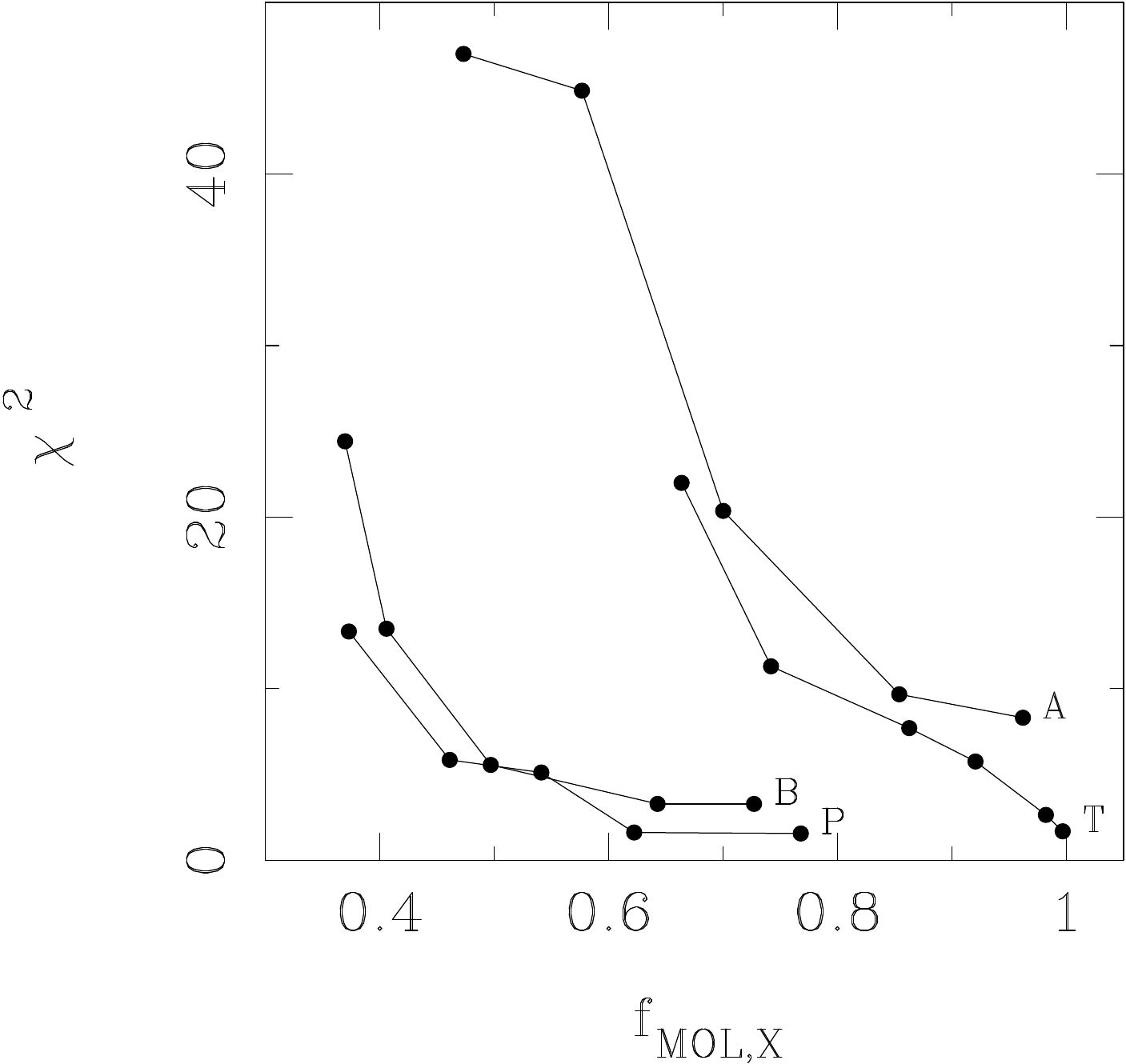}
\caption{Measured $\chi^{2}$ for a single lognormal function that matches the mean and variance of A$_{{\mathrm{V}}}$ for each field selection, plotted versus
the X-dominated fraction, $f_{MOL,X}$.}
\label{fig:fieldchisq}
\end{figure}

\begin{figure}
\includegraphics[width=84mm]{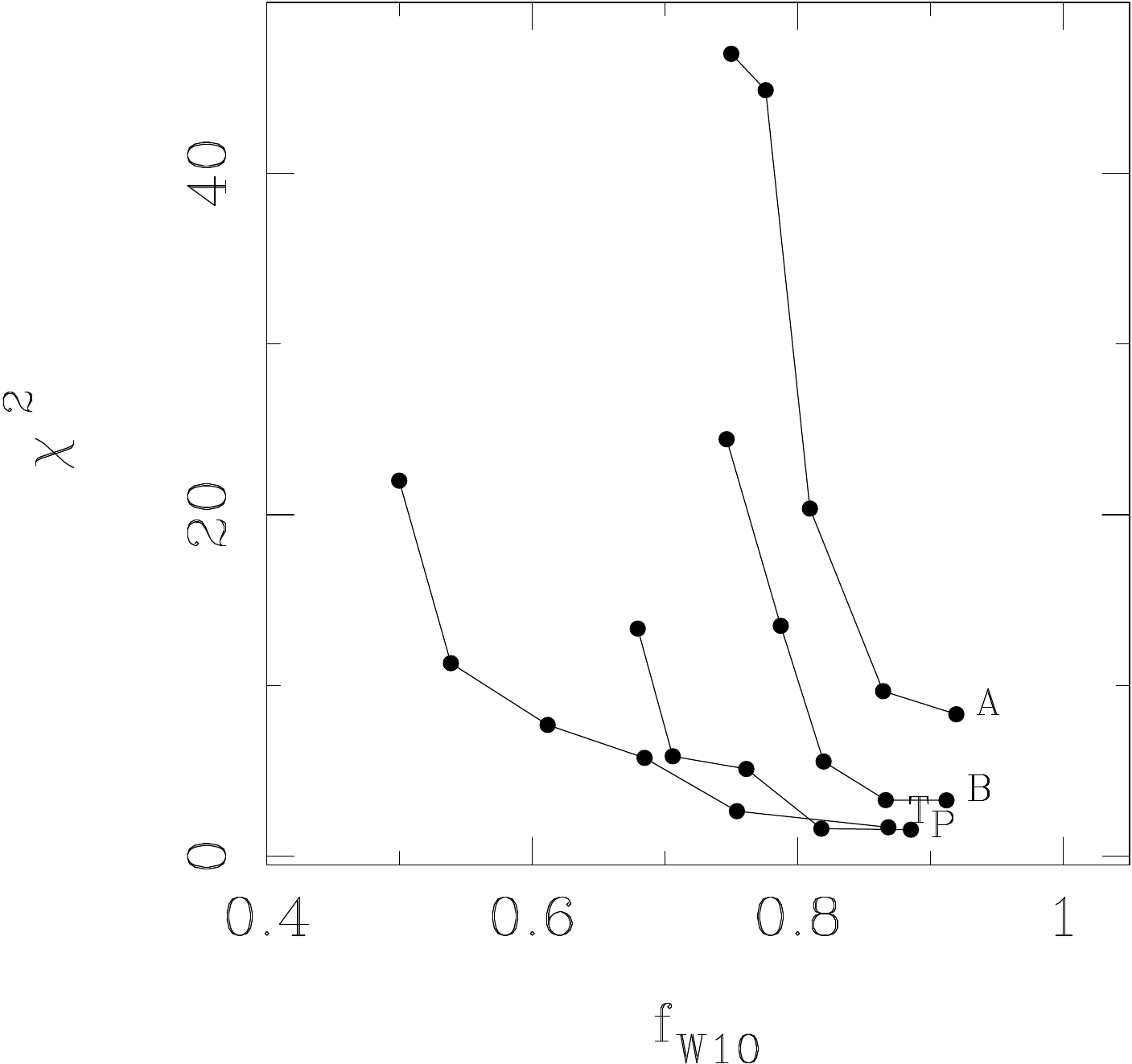}
\caption{Measured $\chi^{2}$ for a single lognormal function that matches the mean and variance of A$_{{\mathrm{V}}}$ for each field selection, plotted versus
$f_{W10}$ = the fraction of CO J=1--0 intensity contributed by pixels exceeding W$_{{\mathrm{CO}}}$~=~10~K~km~s$^{-1}$.}
\label{fig:fw10}
\end{figure}

If the true PDF is not a pure lognormal, $f_{test}$ will not be a good representation of $f_{obs}$.
As a goodness of fit statistic, we use the average of $[\rln(f_{obs,i}/f_{test,i})]^{2} N_{obs,i}$ (= reduced chi-squared, $\chi^{2}$) 
calculated over all PDF bins $i$ where the observed PDF count, $N_{obs,i}$ is greater than 50 (note that the expected error on
$\rln(f_{obs,i}/f_{test,i})$ is $1/\sqrt{N_{obs,i}}$). Field selections which produce a distinct core+tail PDF will therefore 
have a higher $\chi^{2}$ than those that produce a pure lognormal PDF. In the tests below, we do not include accounting for
foreground/background contamination (initially).

To trace the amount of cold (molecular) versus diffuse material incorporated into each field selection, we 
use CO~1--0 data from Dame, Hartmann, \& Thaddeus (2001; DHT01) to infer the contribution to the extinction 
arising from molecular gas traced by CO. We take an X-factor of $1.8{\times}10^{20}$~cm$^{-2}$~(K~km~s$^{-1})^{-1}$ (DHT01)
and a conversion between molecular hydrogen column density and A$_{{\mathrm{V}}}$ of 
$2N_{H_{2}} = 1.8{\times}10^{21}$~cm$^{-2}$~mag.$^{-1}$ (Bohlin, Savage, \& Drake 1978). This results in:
\begin{equation}
\frac{{\mathrm{A}}_{{\mathrm{V,MOL,X}}}}{{\mathrm{mag.}}} = 0.2 \frac{W_{\mathrm{CO}}}{{\mathrm{\; \; K km s}}^{-1}} ,
\end{equation}
where $W_{CO}$ is the integrated intensity of the CO~1--0 line and A$_{{\mathrm{V,MOL,X}}}$ is the inferred 
contribution to the extinction from molecular gas traced linearly by $W_{CO}$.

We convolved the Rowles \& Froebrich A$_{{\mathrm{V}}}$ map to the same resolution as the DHT01 CO map (7.5 arcmin), and resampled
both maps onto the Rowles \& Froebrich grid (2 arcmin). We designated a pixel as ``X-dominated'' if A$_{{\mathrm{V,MOL,X}}}$/A$_{{\mathrm{V}}} \geq 0.5$.
(The label X-dominated here refers to the dominance of material traced linearly by CO via the X-factor, and does not include CO-dark
H$_{2}$ which is probably warmer and more diffuse. Note also that CO emission is used only to define a lower threshold and we do not rely on
W$_{\mathrm{CO}}$ to accurately predict a corresponding A$_{{\mathrm{V}}}$ in the high column density limit where CO may saturate.) 

For each field selection, we determined the fraction of pixels that are X-dominated, and called this $f_{MOL,X}$. For reference, Figure~\ref{fig:fmol}
shows the X-dominated fraction versus spatial scale (in pc) for each cloud. The spatial scales are defined as the square root of the area of the
field selection, using distances of 140~pc (Taurus; Elias 1978), 300~pc (Perseus; Bally et al 2008), and 390~pc (Orion A and B; Mayne \& Naylor 2008).

\begin{figure*}
\includegraphics[width=80mm]{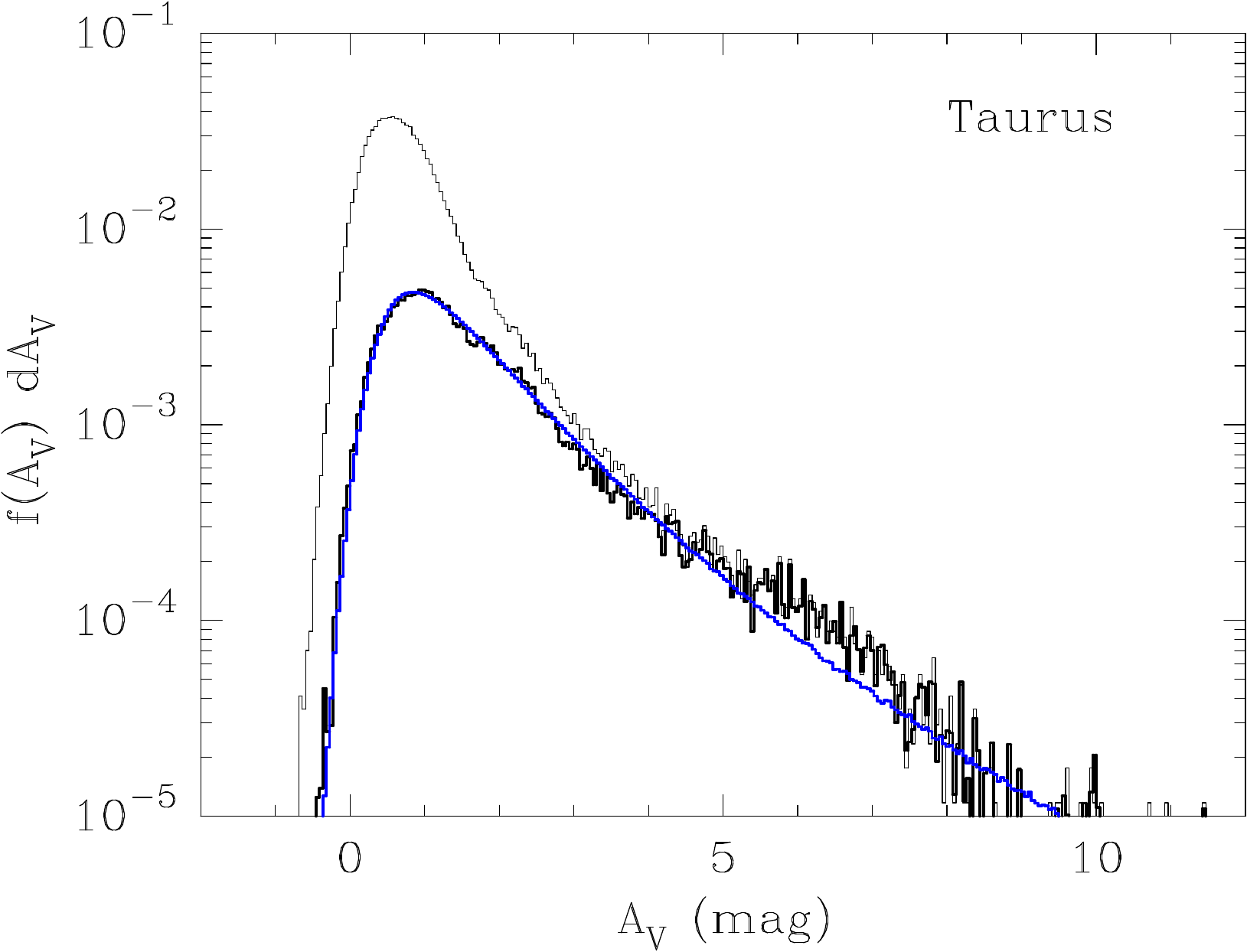}
\includegraphics[width=80mm]{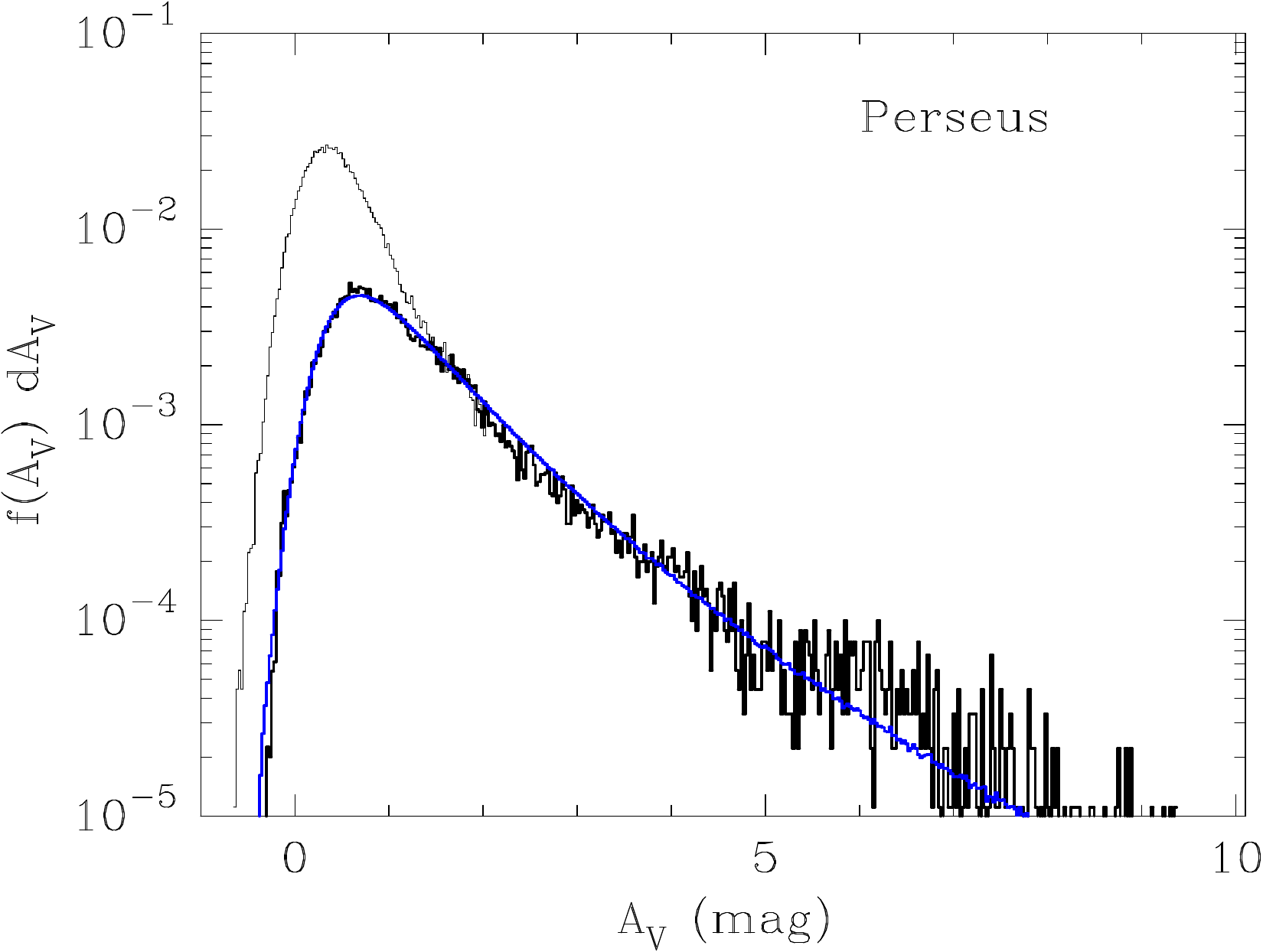}
\includegraphics[width=80mm]{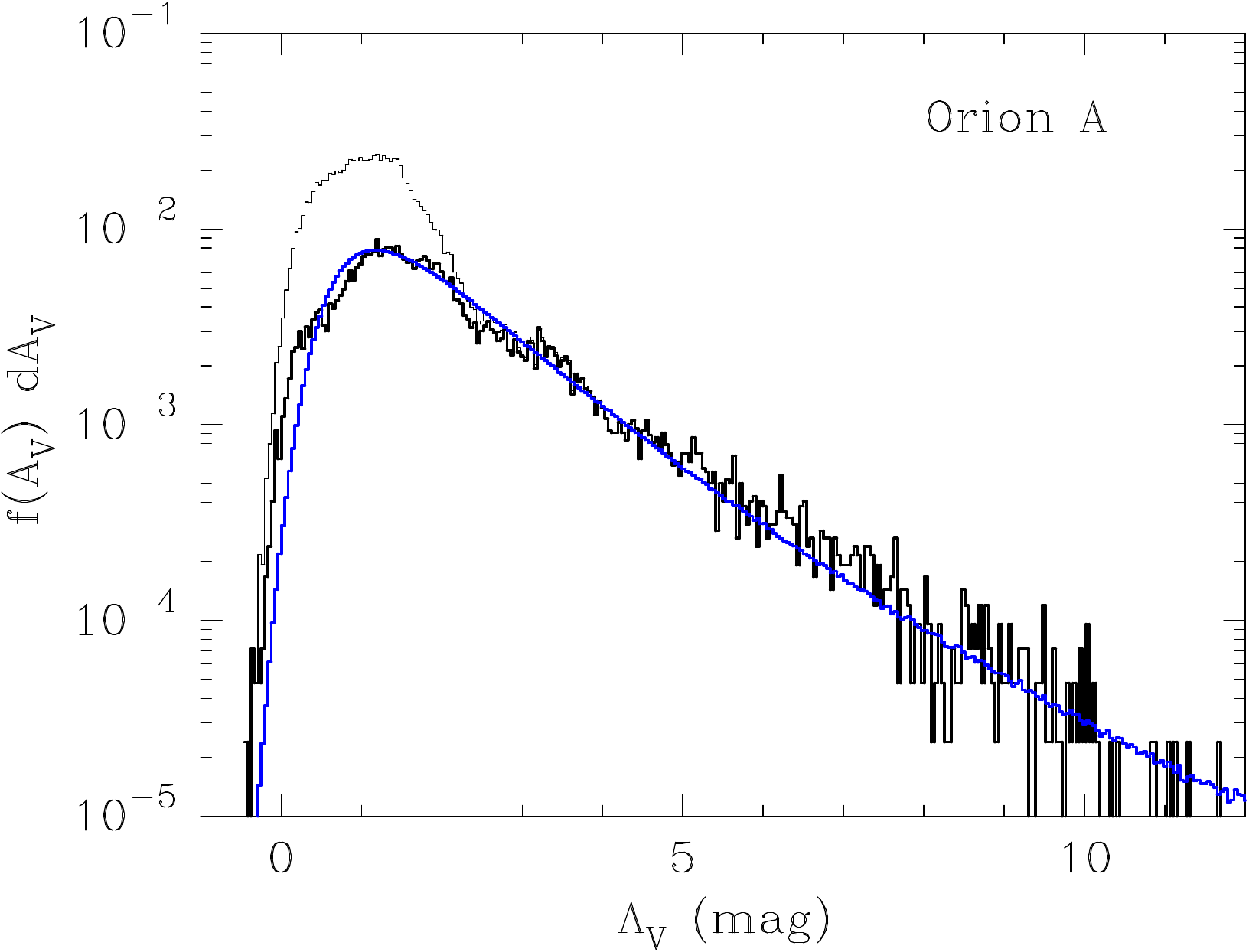}
\includegraphics[width=80mm]{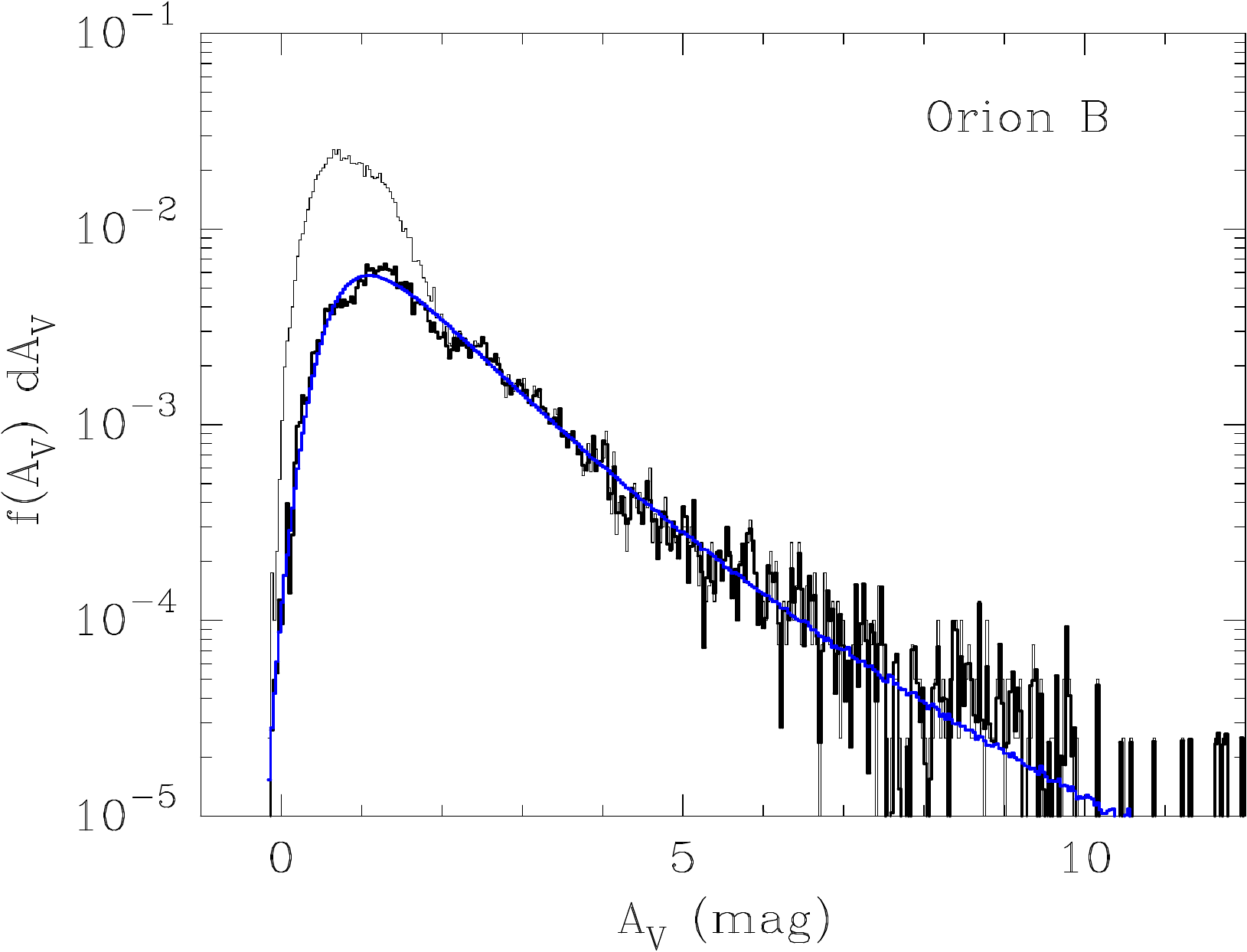}
\caption{Fitted lognormals (blue lines) to the molecule-dominated PDFs (heavy black lines) for
Taurus, Perseus, Orion~A, and Orion~B. Also shown are the wide field PDFs (lighter black lines).}
\label{fig:fieldpdffits}
\end{figure*}

\begin{figure*}
\includegraphics[width=80mm]{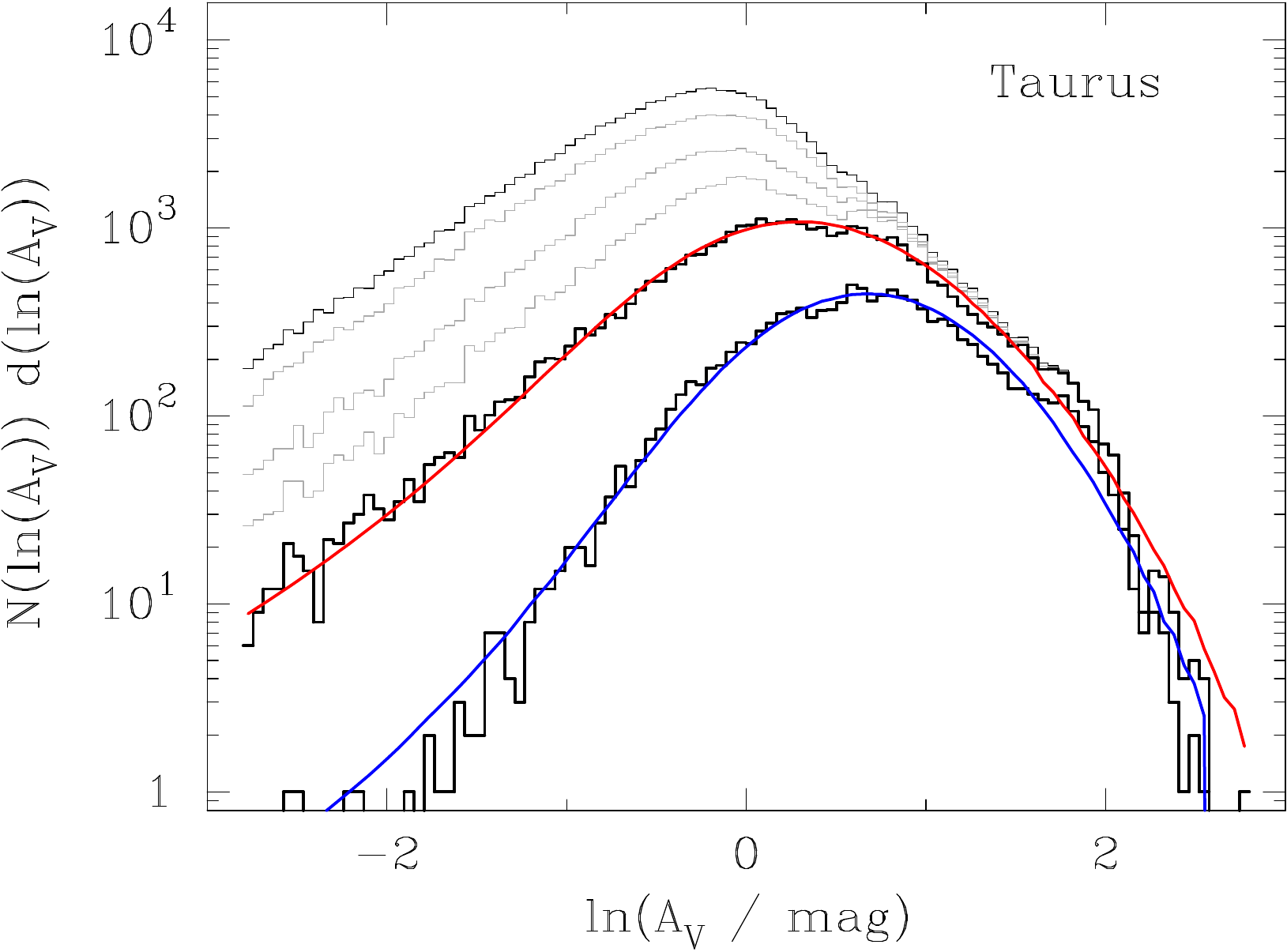}
\includegraphics[width=80mm]{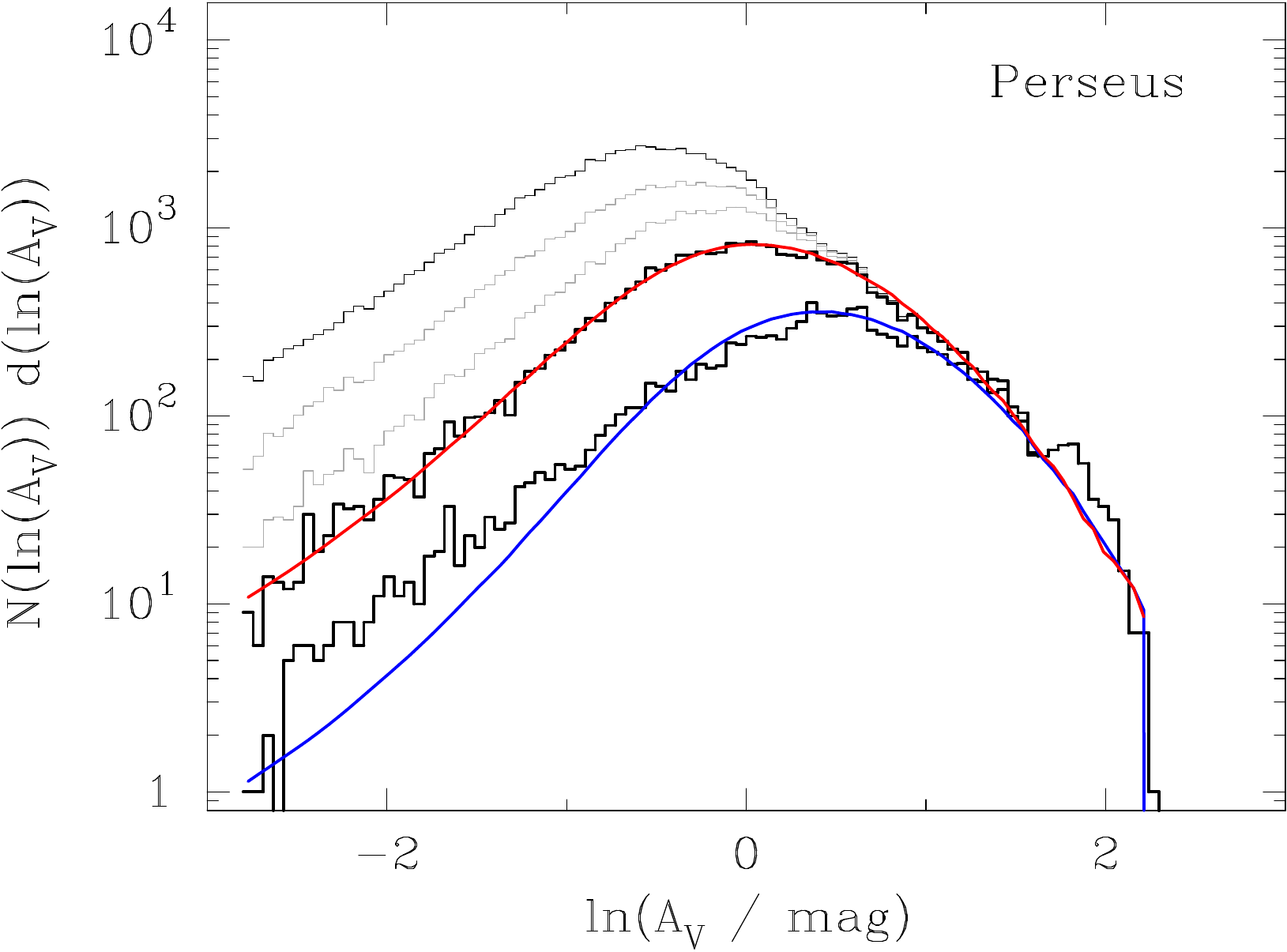}
\includegraphics[width=80mm]{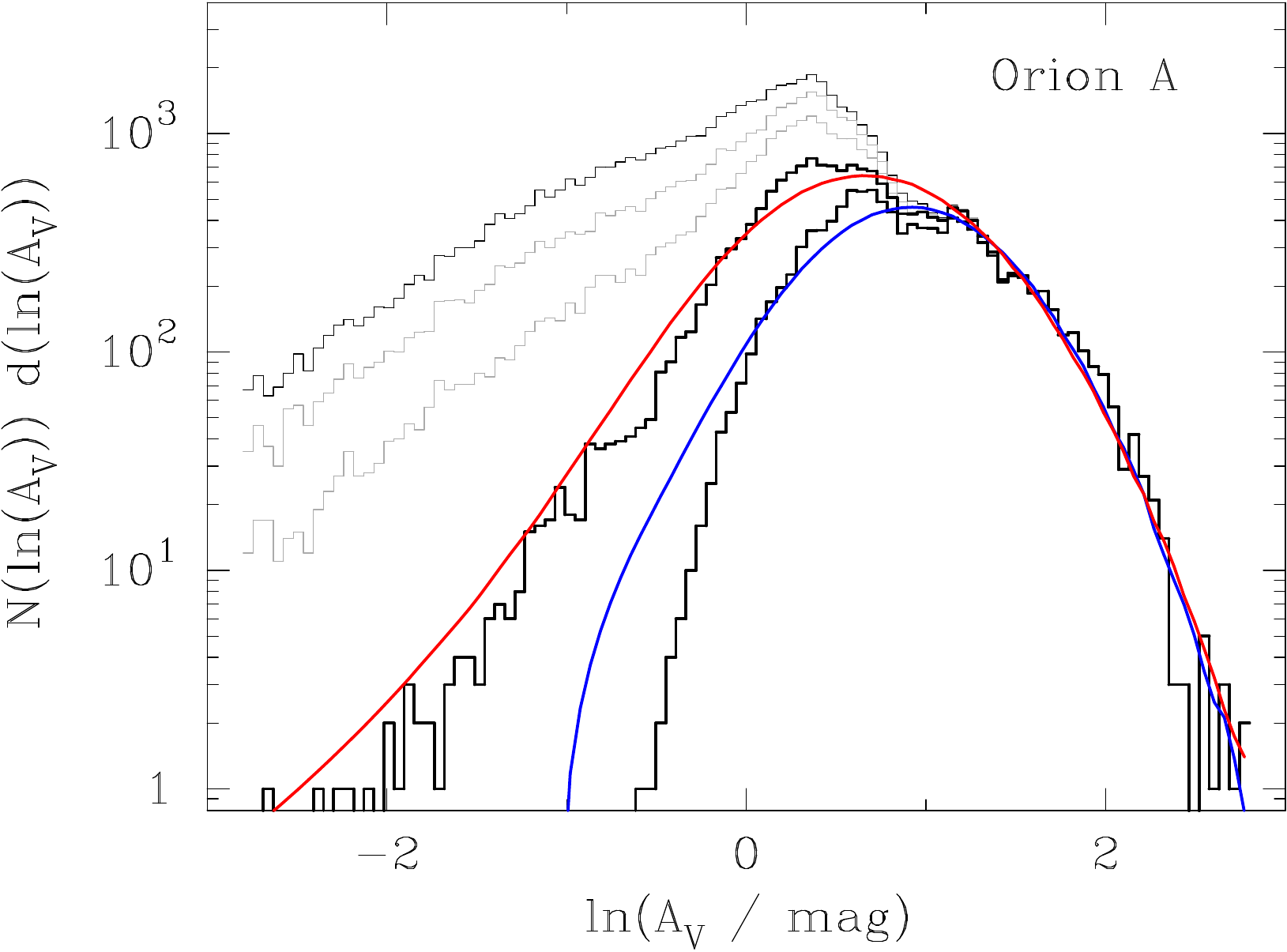}
\includegraphics[width=80mm]{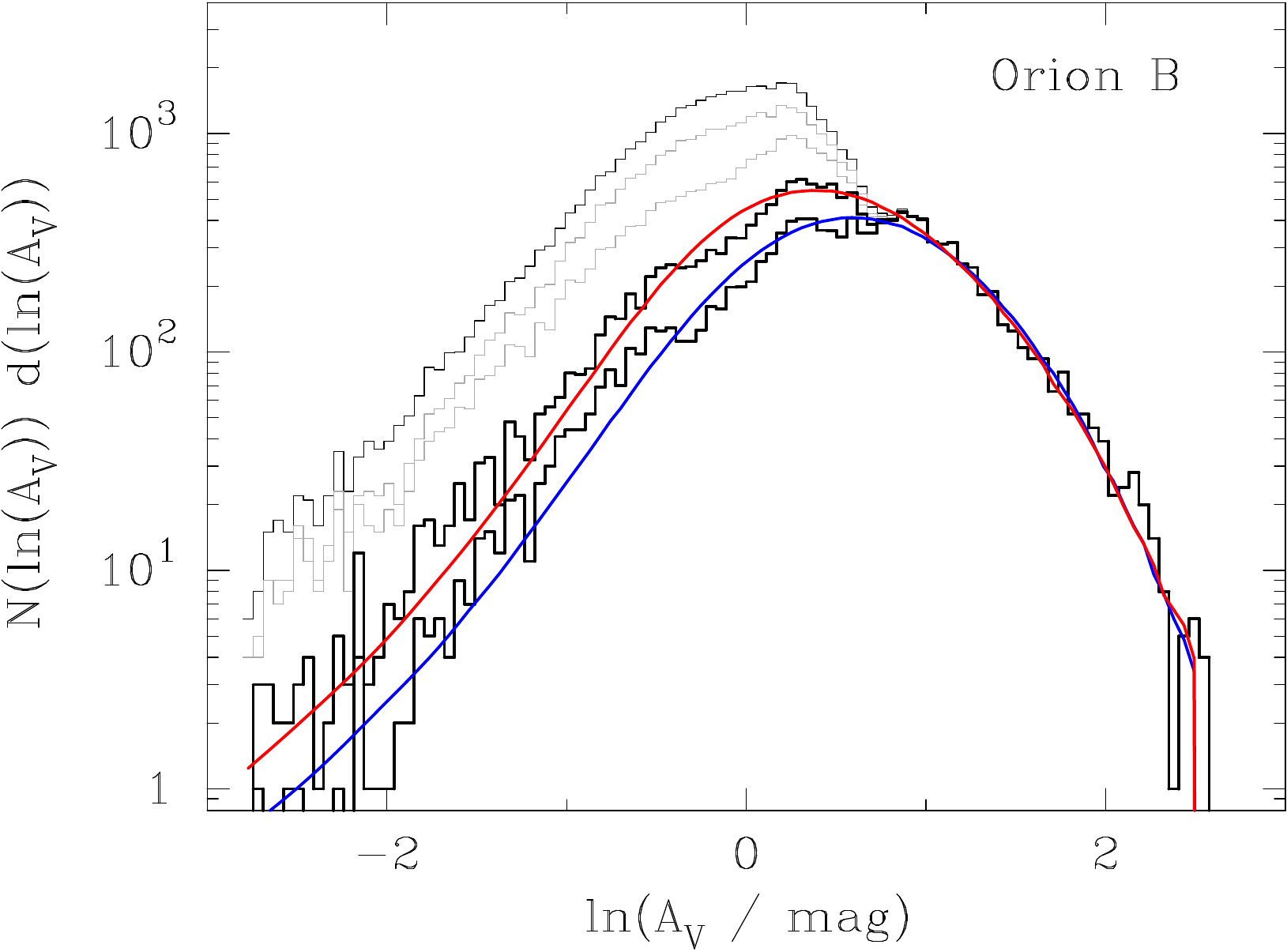}
\caption{Fitted lognormals (blue and red lines) to the PDFs obtained from the two smallest, molecule-rich, field selections 
(heavy black lines) for Taurus, Perseus, Orion~A, and Orion~B. Also shown are the wider field PDFs (lighter black lines) bracketed 
by the largest field selection again shown with a heavier black line.}
\label{fig:fieldpdflogfits}
\end{figure*}

In Figure~\ref{fig:fieldchisq} we plot the measured $\chi^{2}$ (quantifying the success of a single lognormal in representing the observed PDF)
against the measure X-dominated fraction, $f_{MOL,X}$, for each field selection. As the field selection is restricted from wide to
narrow, focusing increasingly on X-dominated extinction structure, a single lognormal function becomes systematically a better fit to
the observed PDF. For some fields (Perseus and Orion B, and Orion A to a lesser extent) the observed $\chi^{2}$ reaches a plateau before the 
narrowest field selection is approached -- i.e. the PDF becomes essentially lognormal and remains lognormal for restricted sub-samples, perhaps
hinting at self-similar behaviour. For reference, the original field selections for Taurus and Perseus in Sections~3 and 4, were at 25~pc ($f_{MOL,X} \sim 0.8$)
and 41~pc ($f_{MOL,X} \sim 0.5$) respectively.
 
There does not appear to be a definitive value of $f_{MOL,X}$ at which the minimum in $\chi^{2}$ is
reached -- a range of $0.6 \lesssim f_{MOL,X} \lesssim 1$ is seen -- but this will depend on the structure of the extinction in the 
immediate vicinity of the main X-dominated region for each cloud. The surroundings of Taurus, for example, are relatively molecule-rich,
by the above definition, compared to the other fields. 
A mass-weighted X-dominated fraction can be calculated, and this suggests that the PDFs 
transition to pure lognormal form when the mass incorporated into the PDF is 80--100\%  X-dominated, by the above definition.

To focus more explicitly on the {\it cold} molecular component, we have also defined a fraction, $f_{W10}$, of sightlines which have
$W_{CO} \geq 10$~K~km~s$^{-1}$. This will more effectively select the coldest $\sim$isothermal gas, while ignoring the warmer
molecular cloud envelopes (e.g. Pineda et al 2010). A plot of $\chi^{2}$ versus $f_{W10}$ is shown in Figure~\ref{fig:fw10}. 
There is a little more coherence to the trend-lines in this Figure (versus Figure~\ref{fig:fieldchisq}), indicating that it is in 
the coldest molecular gas that the PDF tends to a pure lognormal form. A fraction $f_{W10} \gtrsim 0.8$ marks a reasonable
transition point between ``good'' and ``poor'' fits of a single lognormal. The rate at which the fits deteriorate as the cold
molecular zone is exceeded ($f_{W10}$ falls) is variable amongst the clouds however. For Orion A and B, a single lognormal
quickly becomes a very poor fit to the PDF once $f_{W10}$ drops below $\sim$0.8, while for Taurus and Perseus the deterioration
is more gradual. We will return to this point below.

\subsection{Appearance of the PDFs}

In Figure~\ref{fig:fieldpdffits} we show PDFs measured in the molecule-dominated regions for each cloud, compared with the PDF measured in the widest
field selection. For the molecule-dominated PDFs, we have used the largest field selection at which $\chi^{2}$ appears to plateau, and now accounted
for foreground/background contamination. For Orion B, we find that $\mu_{fb} = 0.14$~mag. provides the best fit (which is higher than that in the more 
nearby Perseus and Taurus, for which the previous estimates in Table~1 are good). The extinction structure around Orion A is more complicated: further from
the Galactic plane, the extinctions adjacent to the Orion A X-dominated region are low ($\sim$~0.2~mag.) while closer to the Galactic plane, the 
extinctions are notably higher ($\sim$~1~mag.). The uncertainty in the baseline extinction level can be seen in the
Orion~A PDF at low A$_{{\mathrm{V}}}$: the model lognormal PDF is not a good representation of this bimodal extinction distribution, leading to
relatively higher $\chi^{2}$ for Orion~A than seen in the other fields. We find that foreground/background levels of $\sim$0--0.2~mag. are effectively 
equivalent in terms of $\chi^{2}$ for Orion A, but a more sophisticated baseline level is needed here.  

Figure~\ref{fig:fieldpdffits} shows that the PDF counts are essentially unmodified by the field restriction above a level 
of A$_{{\mathrm{V}}}$~$\sim$~2 (perhaps a little higher for Taurus). The amplitude of the low A$_{{\mathrm{V}}}$ peak for intermediate
field selections lies between the two extremes shown in Figure~\ref{fig:fieldpdffits}.
It is instructive to inspect all the log-space PDFs obtained from the various scale-dependent field selections, and these are
shown in Figure~\ref{fig:fieldpdflogfits}. Red and blue lines show the direct-space lognormal fits translated into the log representation
for the two smallest, molecule-rich, field selections. (As noted above, Orion A has more complicated structure in the surrounding
extinction, and its PDFs are not accounted for very well at low A$_{{\mathrm{V}}}$.) It is evident from Figure~\ref{fig:fieldpdflogfits}
that there are two types of general PDF structure. The PDFs of Taurus and Perseus show more gradual deviation from lognormal behaviour
as the field size is increased to include warmer, diffuse material. In particular, the curvature of the log-log PDFs remains negative (concave
downwards). In contrast, Orion A and B have PDFs with a distinct inflection at $\ln$(A$_{{\mathrm{V}}}$)~$\sim$~0.8--1. This can be
compared to Figure~\ref{fig:fw10} where the PDFs for Orion A and B quickly become non-lognormal once the cold molecular zone
is exceeded by the field selection.

We also extracted PDFs in a number of rectangular windows near the molecule-dominant regions to sample ``diffuse'' PDFs. These are reasonably-well
fitted by lognormals of low dispersion ($\sigma$~$\sim$~0.04--0.45) with the higher dispersions primarily driven by very low mean A$_{{\mathrm{V}}}$.

\section{Discussion - the core/tail duality}

In general, the wide-field PDFs display core+tail structure. The amplitude of the ``core'', and therefore
of any inferred deviations from it, is directly dependent on the rather arbitrary amount of diffuse material included in the 
field selection and therefore incorporated into the PDF. In many cases, the field selection is simply set by the availability of data -- i.e.
determined by the limits over which the cloud is mapped. 

\begin{figure*}
\includegraphics[width=174mm]{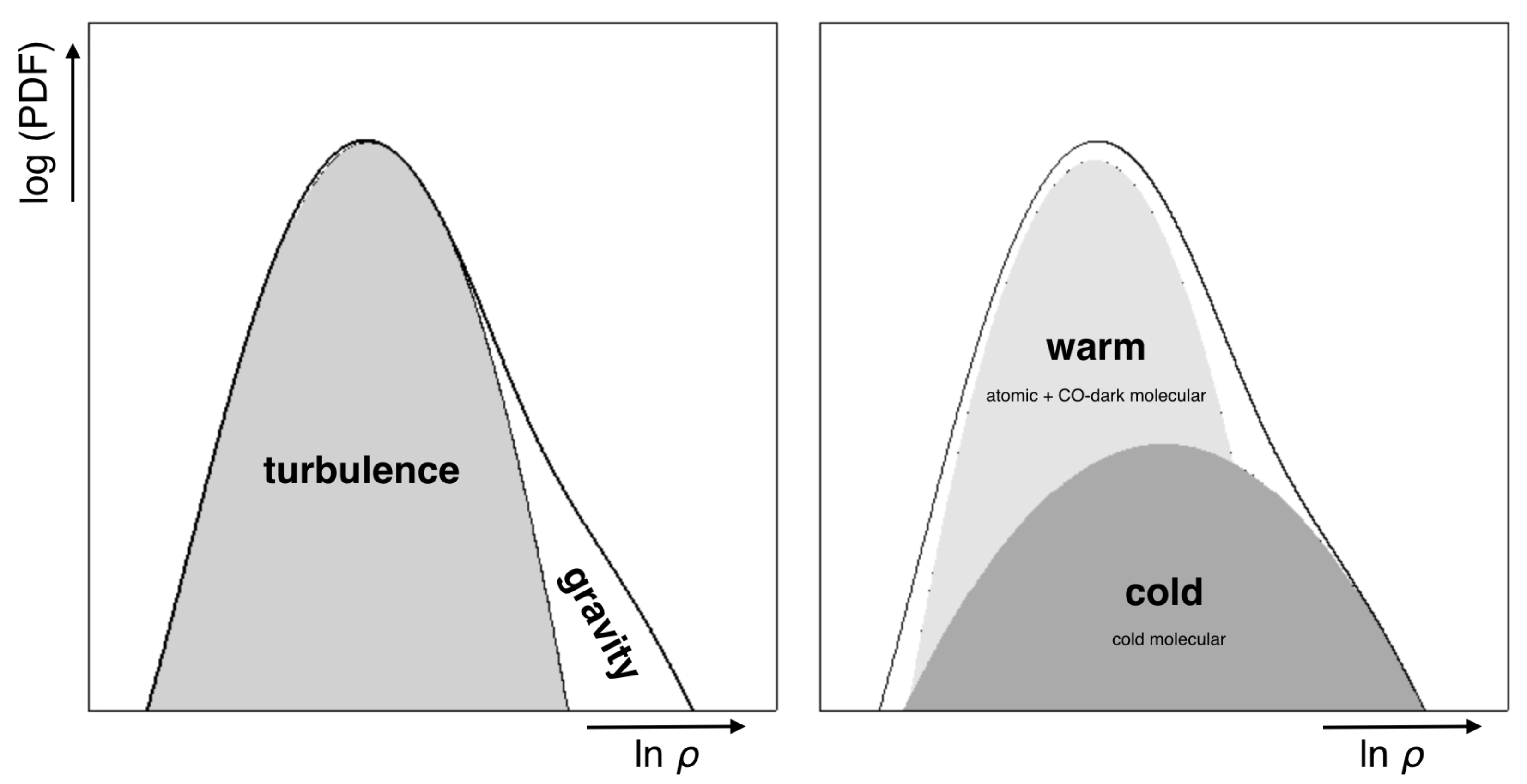}
\caption{Schematic depiction of the core/tail duality for the isothermal turbulence + gravity picture (left) and the multi-phase picture (right).}
\label{fig:coretail}
\end{figure*}

For the molecule-dominated regions, especially the coldest regions, the PDF is
well-described by a pure lognormal, which is the theoretically expected form for isothermal turbulence. It is likely
that the molecule-dominated regions (at least as defined by our CO-based criterion -- which trace the nominal ``molecular clouds'' in each case) 
can be reasonably-well described as isothermal. This means that, at the resolutions probed by the Rowles \& Froebrich data (a few tenths of a pc), 
the core/tail duality is a duality between warm diffuse material and cold molecular material, rather than a duality between nominal 
``lognormal core'' behaviour and gravity-induced power-law behaviour associated with star formation. That is: the ``tail'' part of the wide-field 
PDF is the high extinction part of a lognormal PDF that arises collectively from the main body of the cold molecular cloud. The distinction between these
two scenarios is shown schematically in Figure~\ref{fig:coretail}, though elements of both may play a role in general.

One could ask whether, for an arbitrary field selection, a combination of lognormal core + power-law tail could provide a better description
of the data. But, by Ockham's Razor, there is no need for a power-law contribution, as the observed $\chi^{2}$ 
values are $\sim$unity for a lognormal PDF in the cold, molecule-dominant regions (excepting Orion A for reasons that are understood), 
once a systematic approach to field selection is employed. This is not to say, of course, that gravity {\it definitively} plays no role
in the molecular cloud structure.
 
From the evidence of the PDF alone, it is not possible to distinguish between the turbulence/gravity scenario or the warm/cold scenario, though
in the above we have demonstrated that PDFs sampled from the cold, molecular region are lognormal, and we believe that this favours the
warm/cold picture, with no role for gravity strictly necessary. Kainulainen et al (2011) have also suggested that
the formation of pressure-bound clumps may precede any gravitationally-driven structure.
In a multi-phase turbulent picture, the cold gas will likely have a 
higher Mach number due to its low temperature (e.g. at fixed velocity dispersion), and this would drive a greater dispersion in density 
and therefore column density (the molecular PDFs would have higher log-dispersion than the diffuse PDFs).
It is also known that PDFs generated by the thermal instability will usually be bi-modal, or at least non-lognormal in a similar way to
those observed above (e.g. Audit \& Hennebelle 2010; Gazol \& Kim 2013; Kim, Ostriker \& Kim 2013; Saury et al 2014; Heiner et al 2014, arXiv 1403.6417 submitted),
in part presumably because of the increase in Mach number as the gas cools.

\begin{figure}
\includegraphics[width=84mm]{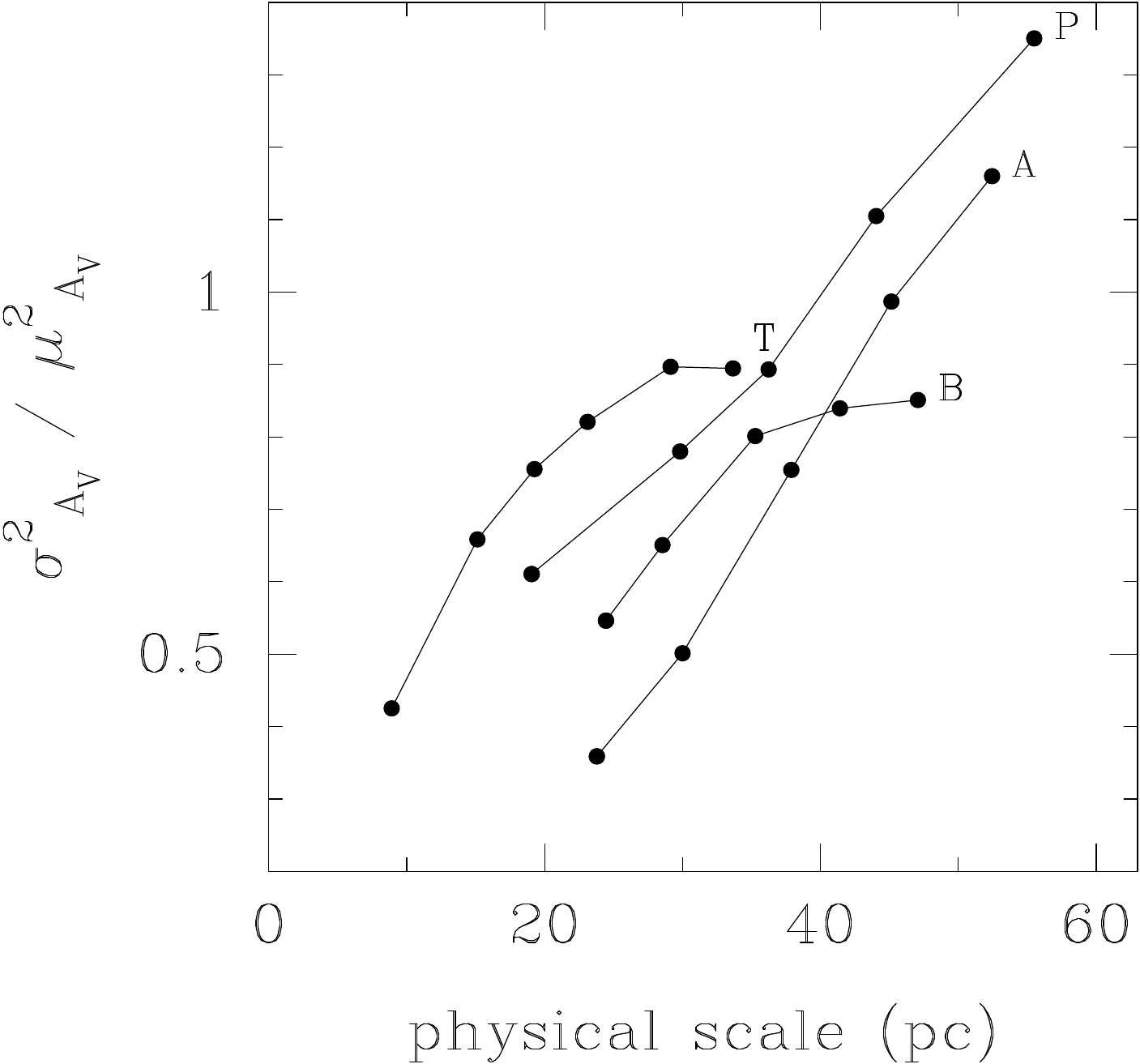}
\caption{Normalised extinction variance, $\sigma^{2}_{{\mathrm{A}}_{{\mathrm{V}}}}/\mu^{2}_{{\mathrm{A}}_{{\mathrm{V}}}}$,
versus spatial scale for the field selections in Taurus (T), Perseus (P), Orion A (A), and Orion B (B).}
\label{fig:fieldvar}
\end{figure}

\begin{figure*}
\includegraphics[width=174mm]{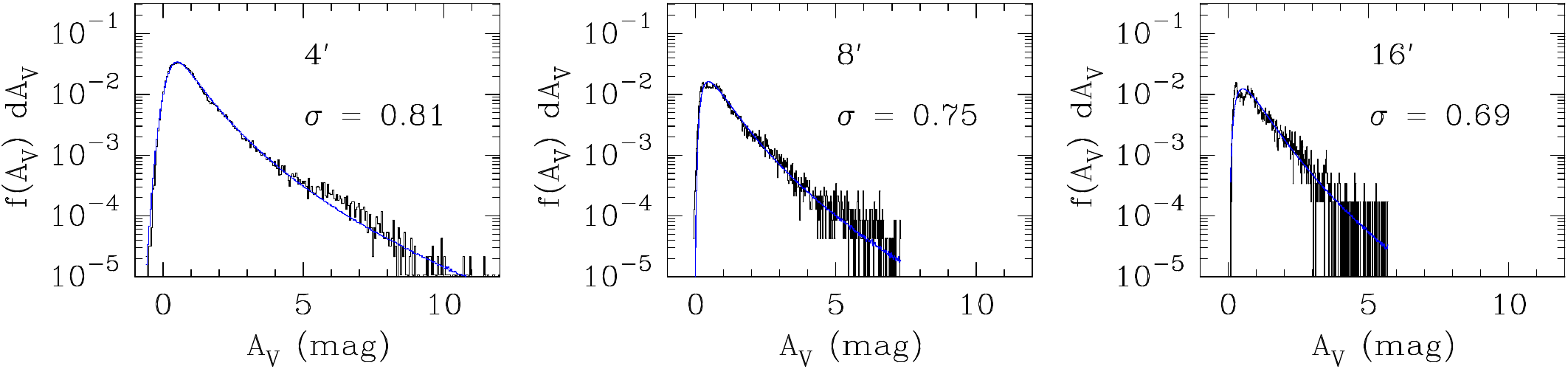}
\caption{PDFs and their dispersions as a function of resolution for Taurus.}
\label{fig:taures}
\end{figure*}

However, since the PDF is a relatively crude measure of the key underlying physics (Tassis et al 2010), this does not rule out a role for gravity in 
the creation of the molecular cloud initially (e.g. Ballesteros-Paredes et al 2011). Neither does it rule out the possible detection of deviations from 
(cold, molecular) lognormality that might be seen at higher resolution. Indeed, for the
highest spatial resolution field (Taurus) there is a slight excess in the PDF, above the nominal lognormal behaviour, near A$_{{\mathrm{V}}} \sim 6-8$~mag.
 (though there does not appear to be a corresponding distinct excess in that A$_{{\mathrm{V}}}$ range in the Taurus 
PDF measured at slightly higher resolution by Kainulainen et al 2009.) Power-law excesses beginning at this A$_{{\mathrm{V}}}$ level have 
in fact been suggested by Froebrich \& Rowles (2010) and Schneider et al (2014). 

If it is the transition between large-scale/warm/diffuse material to smaller-scale/cold/molecular material that drives the change
in PDF structure, then the abruptness of the transition may give clues on the formation mechanism of the cloud. The transitions
in Orion A/B are distinctly more abrupt than those in Taurus/Perseus, as evidenced by the curvature/inflections of the PDFs in log-space
and by the deterioration of the single lognormal fit when the field selection exceeds the cold molecular zone. 
However, for anisotropic clouds, the viewing angle is important (Ballesteros-Paredes et al 2011).

\section{Scale- and resolution-dependence of the column density variance}

It is not our intention here to measure the dispersions of the PDFs, though it is worth a brief comment on the difficulties of this
task. We gave measurements of dispersions in Table~1 solely for the purpose of comparing two methods of measurement. In general, the
measured dispersion will depend on the field selection and the resolution at which the dispersion is measured. We will briefly explore
these issues below.

The lognormal $\sigma$ values for the Orion~A and Orion~B molecule-dominated regions (for the PDFs in Figure~\ref{fig:fieldpdffits}) are 0.53 and 0.71 respectively.
While the variance in A$_{{\mathrm{V}}}$ is higher in Orion~A, the mean A$_{{\mathrm{V}}}$ is also higher by an 
amount that results in a lower $\sigma$ than in Orion~B -- see equation~(\ref{eq:sigdef}). 
Both the Orion $\sigma$ values are notably lower than $\sigma$ listed for Perseus and Taurus in Table~1.
This is in part due to a resolution dependence, as discussed below, but the measured lognormal $\sigma$ in fact 
varies significantly, depending on the field selection. As the field is narrowed more and more, the
mean A$_{{\mathrm{V}}}$ rises, and although the variance also rises, the overall effect is typically to lower the 
measured $\sigma$. This occurs in all the fields analyzed above. To quantify this, we show, in Figure~\ref{fig:fieldvar}, the
variation of normalized extinction variance (related to $\sigma$ via equation~(\ref{eq:sigdef})) as a function of the scale at which it is measured. 
There is no convergence as the molecular zone is breached at small scales.
For the same field shape and centre, the measured $\sigma$ is generally larger for larger field selections -- though at some point a single 
lognormal will cease to represent the PDF. We suggest that a field selection that maximises $\sigma$ subject to simultaneously lying in the
plateau of $\chi^{2}$ may provide an objective definition for $\sigma$ applicable to the cold, molecular zone.

The second issue is that of the resolution-dependence of $\sigma$. Obviously, at higher resolution the field variance must increase
while the mean remains constant. The Taurus field is large enough to perform quantitative tests of this, so we have degraded the resolution
of this field to 8~arcmin and 16~arcmin from the original 4~arcmin. Instead, this can be viewed as moving the cloud to distances of
280~pc and 560~pc respectively. The resulting PDFs and measured $\sigma$ are shown in Figure~\ref{fig:taures}. Interestingly, the 
PDF remains essentially lognormal as the resolution is changed, indicating some form of self-similar behaviour. The measured
$\sigma$ falls as the resolution is degraded. Given the spectral slopes of column density power spectra in molecular clouds,
the dispersion measured in 2D should converge as the resolution is increased, and it may be possible to infer the 
value at convergence. To estimate the corresponding 3D dispersion in density is more difficult, as that is very slowly
convergent, if at all (see discussion in Brunt 2010).

\section{Summary}

We have shown that extinction PDFs of a sample of molecular clouds obtained at a few tenths of a parsec resolution,
probing extinctions up to A$_{{\mathrm{V}}}$~$\sim$~10~magnitudes on scales of a few tenths of a parsec, 
are very well described by lognormal functions provided that the field selection is tightly constrained to the cold, molecular zone and that
noise and foreground contamination are appropriately accounted for. In general, field selections that
incorporate warm, diffuse material in addition to the cold, molecular material will display apparent core+tail PDFs. 
The apparent tail, however, is best understood as the high extinction part of a lognormal PDF arising from 
the cold, molecular part of the cloud.

For a general field selection that includes warm and cold gas, the PDF shows one of two forms. Either the
transition to the molecular zone is abrupt, resulting in an inflection in the log-log PDF structure, or
the transition is more gradual and the log-log PDF retains negative (concave down) curvature at all A$_{{\mathrm{V}}}$.
The former of these is also traced by an abrupt deterioration in $\chi^{2}$ when a single lognormal is used
to fit PDFs that include progressively more diffuse material, while the latter shows a more gradual deterioration
in $\chi^{2}$. These features may provide clues on the formation mechanism of the molecular cloud, but could also
be influenced by the viewing angle.

In terms of fitting the PDF, convolution of the model PDF with the noise PDF is essential. With noise convolution, 
the fitted lognormal PDFs account for both the high A$_{{\mathrm{V}}}$
behaviour of the PDF and the noise-dominated low A$_{{\mathrm{V}}}$ behaviour in the analyzed fields. 
The method used here also has the advantage of directly accounting for both the mean and variance of the
column density, which is desirable. Fitting alternative models with predictable mean and variance,
or use of higher order moments beyond mean and variance, can be simply done.

\section*{Acknowledgments}
C.~B. is funded in part by the UK Science and Technology Facilities Council grant ST/J001627/1 
(``From Molecular Clouds to Exoplanets'') and the ERC grant ERC-2011-StG\_20101014 (``LOCALSTAR''),
both held at the University of Exeter.


\label{lastpage}

\end{document}